\documentclass[preprint]{revtex4}
\usepackage{amsmath}
\usepackage{graphicx}
\usepackage{amssymb}
\usepackage{amsthm, amscd}
\usepackage{amsfonts}
\usepackage{cancel}
\usepackage{times}
\usepackage{pstricks}
\usepackage{wasysym}
\usepackage{ulem}
\usepackage[dvips]{epsfig}
\usepackage{subfigure}
\usepackage{feynmf}

\usepackage{amssymb}

\newcommand{\dslash}{\not{\hbox{\kern-2pt $\partial$}}}
\newcommand{\td}{\tilde} 
\newcommand{\bq}{\begin{equation}} 
\newcommand{\eq}{\end{equation}}
\newcommand{\bqa}{\begin{eqnarray}} 
\newcommand{\eqa}{\end{eqnarray}}
\newcommand{\nn}{\nonumber \\}
\newcommand{\bw}{\begin{widetext}}
\newcommand{\ew}{\end{widetext}}
\newcommand{\tr}{\mbox{tr}}
\begin{document}


\title{Background independent holographic description  : \\ 
From matrix field theory to quantum gravity}

\author{Sung-Sik Lee$^{1,2}$\vspace{0.7cm}\\
{\normalsize{$^1$Department of Physics $\&$ Astronomy, McMaster University,}}\\
{\normalsize{1280 Main St. W., Hamilton ON L8S 4M1, Canada}}\vspace{0.2cm}\\
{\normalsize{$^2$Perimeter Institute for Theoretical Physics,}}\\
{\normalsize{31 Caroline St. N., Waterloo ON N2L 2Y5, Canada}}
}

%

\date{\today}

\begin{abstract}

We propose a {\it local} renormalization group procedure
where length scale is changed in spacetime dependent way.
Combining this scheme with an earlier observation that 
high energy modes in renormalization group play
the role of dynamical sources for 
low energy modes at each scale,
we provide a prescription 
to derive background independent holographic duals
for field theories.
From a first principle construction,
it is shown that the holographic theory 
dual to a $D$-dimensional matrix field theory
is a $(D+1)$-dimensional quantum theory of gravity
coupled with matter fields of various spins.
The gravitational theory has $(D+1)$ first-class constraints
which generate local spacetime transformations in the bulk.
The $(D+1)$-dimensional diffeomorphism invariance is 
a consequence of the freedom to choose different local RG schemes.

\end{abstract}

\maketitle

\section{Introduction}

Can one prove AdS/CFT correspondence\cite{MALDACENA,GUBSER,WITTEN}?
If so, it will not only give us more insight into the precise content of the duality
but also open the door to construct holographic duals for general quantum field theory.
There have been efforts to derive holographic duals directly from boundary field theories\cite{EMIL,DAS,Gopakumar:2004qb,POLCHINSKI09,KOCH,HEEMSKERK10,Faulkner1010,Douglas:2010rc,SUNDRUM,SLEE10,SLEE11,Radicevic,SLEE112}.
One approach is to build up a bulk spacetime, 
by introducing dynamical sources and their conjugate fields 
in exchange of decimating high energy modes 
at each step of renormalization group (RG)\cite{SLEE10,SLEE11,SLEE112}.
This procedure amounts to an exact change of variables
from the original $D$-dimensional fields 
into the $(D+1)$-dimensional variables in functional integration,
where the length scale becomes the extra dimension in the bulk\cite{VERLINDE,LI}.

The first principle construction provides
microscopic justification for 
the dictionary of the AdS/CFT correspondence.
If the prescription is applied to a $D$-dimensional gauge theory,
one obtains a $(D+1)$-dimensional field theory of closed loops 
which are coupled with a two-form gauge field in the bulk\cite{SLEE112}.
The two-form gauge field is an emergent gauge field in the sense that
its dynamics is solely generated by other loop fields.
Because the gauge group is compact, 
there is a topological defect (NS-brane) 
for the gauge field.
Proliferation of the defects describes
quantum tunnelings between different topological sectors.
For a sufficiently large $N$, 
the topological defects are dynamically suppressed,
leading to the deconfinement of the two-form gauge field in the bulk.
It has been emphasized that those phases that admit
`classical' holographic description possess 
a non-trivial quantum order associated with 
the spontaneous suppression of the tunneling between
different topological sectors\cite{SLEE112}.
This is analogous to the quantum order that is present in exotic phases 
of condensed matter systems with an emergent one-form gauge field\cite{Wen_SL},
which often requires a large number of flavors.

Despite some progresses,
the construction\cite{SLEE10,SLEE11,SLEE112} has an important drawback : 
it is not background independent.
As a result, it has not been easy to see the emergence of gravitational theory in the holographic description.
In this paper, we provide a prescription to construct holographic duals 
in a background independent manner.
Using the prescription, we show that 
a $D$-dimensional matrix field theory
can be mapped into a $(D+1)$-dimensional quantum gravity
coupled with matter fields of various spins.
This allows one to identify the boundary field theory
as a quantum theory of gravity in the bulk.

Here is an outline of the paper.
In Sec. II, we start by defining a concrete matrix field theory 
whose holographic dual will be constructed in the remaining of the paper.
A theory is defined by specifying sources for all operators allowed by symmetry.
A local operator is constructed from traces of the fundamental matrix field and its derivatives.
Although the field theory is nominally defined on the flat $D$-dimensional spacetime,
one can view the theory with spacetime dependent sources 
as a theory defined on a curved background spacetime.
%
In Sec. III, we eliminate multi-trace operators 
by introducing a dynamical source and its conjugate field
for each single-trace operator, 
where the conjugate fields represent the operators themselves.
In particular, the $D$-dimensional metric 
and its conjugate field that represents 
the energy-momentum tensor become dynamical.
In Sec. IV, a local coarse graining is performed
where the length scale is increased at a rate 
that depends on spacetime.
As the high energy modes are integrated out,
non-trivial actions are generated 
for the dynamical sources and the conjugate fields.
One contribution is a source dependent determinant 
for the quadratic action of the high energy mode
that is integrated out.
This includes the $D$-dimensional curvature term
of the dynamical metric generated
a la Sakharov's induced gravity\cite{SAKHAROV}.
The other contribution represents 
double-trace operators generated from quantum correction,
which become a quadratic action for the conjugate field
as the double-trace operators
are removed by another set of auxiliary fields.
In Sec. V, we take the advantage of 
the $D$-dimensional diffeomorphism invariance,
which is ensured by the fact that the $D$-dimensional metric is fully dynamical,
to shift the $D$-dimensional coordinates of the low energy modes
relative to the coordinates of the high energy field.
In Sec. VI, it is shown that one can construct a $(D+1)$-dimensional
theory as the coarse graining procedure is repeatedly applied 
to the low energy mode.
In Sec. VII, we show that the theory in the bulk
takes the form of a $(D+1)$-dimensional canonical quantum gravity
if one interprets the extra dimension associated with the scale 
as a time.
In particular, there are $(D+1)$ local constraints which originate
from the fact that the partition function is independent 
of the local RG scheme :
the partition function is invariant under the changes
in the speed of local coarse graining and the $D$-dimensional shift.
From this, it can be shown that those constraints are first-class,
which generate $(D+1)$-dimensional local spacetime transformations.
In Sec. VIII, we apply the prescription 
to a simple toy model ($0$-dimensional matrix model)
to illustrate the main idea in the simplest setting.
In Sec. IX, the difference between 
the present holographic description 
and the conventional RG
is contrasted.
In particular, we emphasize the fact that
the beta function is promoted to 
a `Heiserberg' equation for {\it quantum } operators
in the holographic description.

\section{Model}
\subsection{Matrix field theory}
Consider a matrix quantum field theory defined on a $D$-dimensional flat spacetime.
To be concrete, we consider a theory 
of $N \times N$ real traceless symmetric matrix field $\Phi(x)$
with the global $O(N)$ symmetry
under which the matrix field transforms as an adjoint field.
The `partition function' is 
\bqa
Z[ {\cal J}] & = & \int D \Phi ~~ \exp \left[ i N^2 \int d^{D} x ~ \left( -{\cal J}^{m} O_{m} + V [ O_{m }; {\cal J}^{ \{ m_i \}, \{  \nu^i_j \} } ] \right) \right].
\label{Z}
\eqa
Here $O_m$'s denote single-trace operators constructed from $\Phi$ and its derivatives. 
In general, one can take $\{ O_m \}$ to be a complete set of primary single-trace operators.
Here we use the basis where $O_m$ takes the form of
\bqa
O_{[q+1;\{\mu^i_j\}]}  = \frac{1}{N} \tr \left[
\Phi
\left( \partial_{\mu^1_1} \partial_{\mu^1_2} .. \partial_{\mu^1_{p_1}} \Phi \right)
\left( \partial_{\mu^2_1} \partial_{\mu^2_2} .. \partial_{\mu^2_{p_2}} \Phi \right)
...
\left( \partial_{\mu^{q}_1} \partial_{\mu^{q}_2} .. \partial_{\mu^{q}_{p_q}} \Phi \right)
\right],
\label{o}
\eqa
where $q+1$ is the order in the matrix field,
and $\{ \mu^i_j \}$ specifies the spacetime indices.
General single-trace operators can be written as linear combinations of 
these operators and their derivatives.
For simplicity, we assume that there is no boundary in spacetime.
Any operator that has overall derivatives is removed by integration by part in Eq. (\ref{Z}).
Throughout the paper, we will use the compressed label, say $m$ to denote the full indices, $[q, \{ \mu^i_j\}]$ of a single-trace operator.
Explicit indices will be used only when it is needed.
${\cal J}^m(x)$ is the spacetime dependent sources for the corresponding operator $O_m$.
The information on the signature of the background metric is solely encoded in the sources.
We assume that the spacetime has the Minkowskian metric with the signature $(-1,1,1,..,1)$ for $x^\mu$ with $\mu=0,1,..,(D-1)$.
$V$ represents a multi-trace deformation,
\bqa
V[ O_{m }; {\cal J}^{ \{ m_i \}, \{  \nu^i_j \} } ] & = & \sum_{q=1}^\infty 
 {\cal J}^{ \{ m_i \}, \{  \nu^i_j \} } 
O_{m_1}
\left( \partial_{\nu^1_1} .. \partial_{\nu^1_{p_1}} O_{m_2} \right)
\left( \partial_{\nu^2_1}  .. \partial_{\nu^2_{p_2}} O_{m_3} \right)
...
\left( \partial_{\nu^{q}_1}.. \partial_{\nu^{q}_{p_q}} O_{m_{q+1} } \right), \nn
\label{V}
\eqa 
where $ {\cal J}^{ \{ m_i \}, \{  \nu^i_j \} }(x)$'s are sources for multi-trace operators.
All repeated indices are summed over.

To make sense of the partition function,
the theory should be regularized.
Here we use the Pauli-Villar regularization.
Namely, the sources for high derivative terms are turned
on in the quadratic action for the matrix field to suppress UV divergence
in loop integrals.
For example, one can use a regularized kinetic term,
$-\tr[ \Phi \Box e^{ -\frac{\Box}{M^2} } \Phi ]$,
where $\Box = \partial_\mu \partial^\mu$.
The mass scale $M$ in the higher derivative terms
plays the role of a UV cut-off.
It is noted that the divergence in the determinant of the quadratic action
is not regularized by the higher derivative terms.
In this sense, the partition function itself is not well defined.
What is well defined is the ratio between two partition functions
with two different sets of sources
where the divergences from the determinants cancel.
For example, the divergence in the determinant is canceled
in correlation functions of local operators.

\subsection{From flat to curved background spacetimes}
Suppose the manifold is endowed with a background metric $G_{\mu \nu}$.
One can define covariant operators that transform as tensor density of weight one 
under coordinate transformations,
\bqa
O^{G}_{n } =  
\frac{1}{N} \sqrt{|G|}~
tr \left[
\Phi
\left( \nabla_{\mu^1_1}^G \nabla_{\mu^1_2}^G .. \nabla_{\mu^1_{p_1}}^G \Phi \right)
\left( \nabla_{\mu^2_1}^G \nabla_{\mu^2_2}^G .. \nabla_{\mu^2_{p_2}}^G \Phi \right)
...
\left( \nabla_{\mu^{q}_1}^G \nabla_{\mu^{q}_2}^G .. \nabla_{\mu^{q}_{p_q}}^G \Phi \right)
\right],
\eqa
where $\sqrt{|G|} \equiv  \sqrt{ |\det G_{\alpha \beta}|}$ 
and $\nabla_\mu^G$ is the covariant derivative 
associated with the background metric.
Any operator $O_m$ defined on the flat spacetime 
can be expressed as a linear combination of the covariant operators,
\bqa
O_{m }(x) & = & c^{~~n }_{m }(G) O^{G}_{n }(x), 
\eqa
where $c^{~~n }_{m }(G)$ is the transformation matrix.
It is a function of the metric $G_{\mu\nu}$ and its derivative
at the position $x$.
For example,
\bqa
O_{[2,\mu\nu]} & = & 
\frac{1}{\sqrt{|G|}} \left[
O^{G}_{[2,\mu\nu]}
+ \Gamma^\lambda_{\mu \nu}
O^{G}_{[2,\lambda]}
\right], 
\eqa
where
$O_{[2,\mu\nu]} = \frac{1}{N} \tr ( \Phi \partial_\mu \partial_\nu \Phi )$,
$O^G_{[2,\mu\nu]} = \frac{1}{N} \sqrt{|G|} \tr ( \Phi \nabla^G_\mu \nabla^G_\nu \Phi )$,
$O^G_{[2,\lambda]} = \frac{1}{N} \sqrt{|G|} \tr ( \Phi \nabla^G_\lambda \Phi )$,
and
$\Gamma^\lambda_{\mu \nu}$ is the Christoffel symbol for the metric $G_{\mu \nu}$.
Therefore, the same Lagrangian in Eq. (\ref{Z}) can be written in terms of these `covariant operators',
\bqa
{\cal L} =  N^2 \Bigl\{
-{\cal J}^{G; m } O^G_{m} + V [ O^G_{m}; {\cal J}^{G;  \{ m_i \}, \{  \nu^i_j \}  } ] \Bigr\}, 
\eqa
where
\bqa
{\cal J}^{G;m }(x) & = &  {\cal J}^{n }(x) c^{~~m }_{n }(G).
\eqa
The inverse of the transformation is given by
\bqa
O^{G}_{m }(x) & = & d^{~~n }_{m }(G) O_{n }(x), \nn
{\cal J}^{m }(x) & = &  {\cal J}^{G;n }(x) d^{~~m }_{n }(G), 
\eqa
with $d_{a }^{~~m }(G) c_{m }^{~~b }(G) = \delta_a^{~b}$.
The multi-trace operators can be also expressed in terms of 
the covariant operators and the covariant derivatives of them,
\bqa 
V[ O^G_{m }; {\cal J}^{G; \{ m_i \}, \{  \nu^i_j \} } ] & = & \sum_{q=1 }^\infty
\frac{ {\cal J}^{G; \{ m_i \}, \{  \nu^i_j \} } }{ |G|^{\frac{q}{2}}} \nn
&& 
O^G_{m_1}
\left( \nabla_{\nu^1_1}^G .. \nabla_{\nu^1_{p_1}}^G O^G_{m_2} \right)
\left( \nabla_{\nu^2_1}^G  .. \nabla_{\nu^2_{p_2}}^G O^G_{m_3} \right)
...
\left( \nabla_{\nu^{q}_1}^G .. \nabla_{\nu^{q}_{p_q}}^G O^G_{m_{q+1}} \right),
\label{V2}
\eqa 
where ${\cal J}^{G; \{ m_i \}, \{  \nu^i_j \}}$ can be similarly
expressed as a linear combination of ${\cal J}^{\{ m_i \}, \{  \nu^i_j \} } $
so that Eq. (\ref{V2}) coincides with Eq. (\ref{V}).
The explicit form of the transformation is not important.
The factor of $|G|^{-\frac{q}{2}}$ is introduced
to make the whole expression to have weight one
when ${\cal J}^{G; \{ m_i \}, \{  \nu^i_j \} } $ has weight zero.

This means that 
the original theory defined on the flat spacetime can be viewed 
as a theory defined on a curved spacetime with any background metric.
The theory does not depend on the background metric because
different choices of metric can be compensated by 
metric dependent sources.
However, there is a natural choice of metric.
We choose the metric $G^{(0) \alpha \beta}$, 
such that the two derivative kinetic term 
takes the canonical form, 
that is, $J^{(0) [2,\mu \nu]} = G^{(0) \mu \nu}$, 
where $J^{(0) [2,\mu \nu]}$ denotes the source for
$ \sqrt{|G^{(0)}|} tr \left[ \Phi \nabla_\mu^{G^{(0)}} \nabla_\nu^{G^{(0)}} \Phi  \right]$.
This is always possible 
because one can make $J^{(0) [2,\mu \nu]}$ symmetric in $\mu$ and $\nu$ without loss of generality.
In $D>2$, there is a unique metric that satisfies 
the canonical condition, $J^{(0) [2,\mu \nu]} = G^{(0) \mu \nu}$
for a given set of sources (For a proof of this, see Appendix A).
Such choice of metric is not unique at $D=2$ 
where one needs an extra condition to fix 
the freedom associated with the dilatation.
Here we assume that $D>2$.
In this choice of the background metric, 
the kinetic term takes the canonical form
\bqa
{\cal L } & = & 
- N \sqrt{|G^{(0)}|} ~
G^{(0)\mu \nu} 
tr \left[ \Phi \nabla_\mu^{G^{(0)}} \nabla_\nu^{G^{(0)}} \Phi  \right] 
+ ...,
\label{bm}
\eqa
where the same metric is used for the covariant derivative in each tensorial operator
and the source for the kinetic term with two derivatives.
In this sense, an action on a flat spacetime 
with spacetime dependent sources 
defines a natural curved background spacetime.
Physically, this amounts to measuring the distance on the manifold based on the cost of the action in the limit that the amplitude of field is small and the field changes slowly in spacetime.
For example, one can set the distance between two points to be $1$ 
when the quadratic action that is needed to twist fields between the two points 
is $N$ per unit twist and per unit square modulus for each field.
In this choice of metric, we denote
\bqa
{\cal J}^{(0) m }(x) & \equiv &  {\cal J}^{n}(x) c^{~~m }_{n }(G^{(0)}), \nn
O^{(0)}_{m }(x) & \equiv & d^{~~n }_{m }(G^{(0)}) O_{n}(x)
\eqa
to write
\bqa
{\cal L} = N^2 \Bigl\{ -{\cal J}^{(0) m } O^{(0)}_{m} + V [ O^{(0)}_{m}; {\cal J}^{(0); \{ m_i \}, \{ \nu^i_j \} } ] \Bigr\}. 
\label{13}
\eqa
Here the background geometry is in general curved, but it is non-dynamical.

\section{Auxiliary fields and Gauge fixing}
In the holographic construction\cite{SLEE10,SLEE11,SLEE112}, 
sources become dynamical in the bulk.
Therefore one needs to introduce a dynamical field in the bulk
for each independent operator.
Since multi-trace operators can be written 
as products of single-trace operators,
it is convenient to remove the multi-trace operators at the expense of making the sources
for the single trace operators dynamical.
We introduce a pair of auxiliary fields for every single-trace operator\cite{SLEE112},
\bqa
Z & = & \int D  j^{(1) n} D p^{(1)}_n  D  \Phi ~~ 
e^{i \int d^{D} x ~ {\cal L}_1 },
\eqa
where
\bqa
{\cal L}_1 
& = & N^2 \Bigl\{  j^{(1) m } ( p_{m }^{(1)} - O^{g}_{m } )
- {\cal J}^{(0)m } f_m^{~~n}(G^{(0)},g) p_{n }^{(1)}  + 
V[ f_m^{~~n}(G^{(0)},g) p_{n }^{(1)}; {\cal J}^{(0); \{ m_i \}, \{ \nu^i_j \} }] \Bigr\} \nn
 & = &  N^2 \Bigl\{  \left( j^{(1) n } - {\cal J}^{(0)m } f_m^{~~n}(G^{(0)},g) \right) p_{n}^{(1)}
 - j^{(1) m } O^{g}_{m } + 
V[ f_m^{~~n}(G^{(0)},g) p_{n }^{(1)}; {\cal J}^{(0); \{ m_i \}, \{ \nu^i_j \} }] \Bigr\}. 
\label{L1}
\nn
\eqa
Here $j^{(1)n}(x)$ and $p^{(1)}_n(x)$ are $D$-dimensional auxiliary fields
which in general carry $D$-dimensional spacetime indices.
$g_{\mu \nu}$ is an arbitrary $D$-dimensional metric that is used to define 
a new set of tensorial operators $O^g_m$.
It is noted that $g$ is in general different from $G^{(0)}$.
The matrix that transforms fields 
defined with different background metrics 
is given by
\bqa
f_m^{~~n}(G^{(0)},g) & = &  d_m^{~~a}(G^{(0)}) c_a^{~~n}( g ),
\eqa
which satisfies 
$f(g,g)= I$, 
$f(g^{(1)},g^{(2)}) f(g^{(2)},g^{(3)}) = f(g^{(1)},g^{(3)})$ 
and 
$f(g^{(1)},g^{(2)}) f(g^{(2)},g^{(1)}) = I$. 
$j^{(1)n}$'s are dynamical sources for single-trace operators,
and $p^{(1)}_n$'s represent the operators themselves\cite{SLEE112}.
In Eq. (\ref{L1}), $j^{(1)n}$ is a Lagrangian multiplier 
that enforces the constraint $p^{(1)}_n = O^g_n$.
Once $j^{(1)n}$ and $p^{(1)}_n$ are integrated out,
Eq. (\ref{L1}) becomes Eq. (\ref{13}).

Since the partition function is independent of the metric $g_{\mu \nu}$,
we can formally integrate over different choices of $g_{\mu \nu}$ and divide by the volume of the space of metric (gauge volume),
\bqa
Z & = & \int \frac{Dg D j^{(1)n} D p^{(1)}_n  D  \Phi}{V_{gauge}} ~~ 
e^{i \int d^{D} x ~ {\cal L}_1 }.
\eqa
The resulting theory has a gauge symmetry associated with different choices of the metric $g_{\mu \nu}$.
It is emphasized that this gauge redundancy is different from the coordinate redundancy generated by the $D$-dimensional diffeomorphism.
Rather it is associated with different choices of background metric
in a fixed coordinate system.
Under the gauge transformation, the fields transform as
\bqa
g_{\mu \nu} & \rightarrow & g_{\mu \nu}^{'} = 
g_{\mu \nu} + \delta g_{\mu \nu}, \nn
j^{(1)m} & \rightarrow & j^{(1)m'} = j^{(1)n} f_n^{~m}( g^{}, g + \delta g), \nn
p^{(1)}_m & \rightarrow & p^{(1)'}_m = f_m^{~n}( g + \delta g, g^{}) p^{(1)}_n, \nn
O^{g}_m & \rightarrow & O^{g+\delta g}_m = f_m^{~n}( g + \delta g, g^{}) O^{g}_n.
\eqa
Again we fix the gauge by requiring that the quadratic kinetic term has the canonical form.
Namely, we will choose the gauge where
$j^{(1)[2,\mu \nu]'} = g^{\mu \nu '}$.
Here we have to take into account a non-trivial determinant
in gauge fixing because the sources are dynamical
unlike the case in Sec. II. 
The determinant associated with the gauge fixing can be obtained 
from the standard Fadeev-Popov method.
We first define $\Delta(j,g)$ such that 
\bqa
\int D \delta g ~~ \delta \Bigl( 
 j^{n} f_n^{~[2,\mu \nu]}( g, g + \delta g) 
- (g+\delta g)^{\mu \nu}
\Bigr) \Delta(j,g) = 1.
\eqa
This identity is inserted into the partition function,
\bqa
Z & = & \int \frac{Dg D j^{(1)n} D p^{(1)}_n  D  \Phi D \delta g}{V_{gauge}} ~~  \nn
&& ~~ 
\Delta(j^{(1)},g) 
\delta \left( 
 j^{(1)n} f_n^{~[2,\mu \nu]}( g^{}, g^{} + \delta g) 
- (g^{}+\delta g)^{\mu \nu}
\right)
e^{i \int d^{D} x ~ {\cal L}_1[j^{(1)},p^{(1)},g^{\mu \nu},\Phi] }. 
\eqa
By changing the variables,
\bqa
G^{(1)\mu \nu} & = & g^{\mu \nu} + \delta g^{\mu \nu}, \nn
J^{(1)m} & = & j^{(1)n} f_n^{~m}( g^{}, g^{} + \delta g), \nn
P^{(1)}_m & = & f_m^{~n}( g^{} + \delta g, g^{}) p^{(1)}_n,
\eqa
and by using the gauge invariance of ${\cal L}_1$ and the facts that
\bqa
\left| \frac{\partial j^{(1)}}{\partial J^{(1)}} \right|
 \left| \frac{\partial p^{(1)}}{\partial P^{(1)}} \right| &=& 1, \nn
\left| \frac{\partial g^{}}{\partial G^{(1)}} \right| &=& 1, \nn
\Delta( j^{(1)n} f_n^{~m}( g^{}, g^{} + \delta g), g^{} + \delta g ) & = & 
\Delta( j^{(1)m}, g^{} ),
\eqa
we obtain
\bqa
Z & = & \int \frac{DG^{(1)} D  J^{(1) n} D P^{(1)}_n  D  \Phi D \delta g}{V_{gauge}} ~~ 
\Delta(J^{(1)},G^{(1)}) 
\delta \left( 
 J^{(1)~[2,\mu \nu]} 
- G^{(1) \mu \nu}
\right)
e^{i \int d^{D} x ~ {\cal L}_1^{'}[J^{(1)},P^{(1)},G^{(1) \mu \nu},\Phi] } \nn
& = & \int D J^{(1)n} D P^{(1)}_n  D  \Phi  ~~ 
\Delta(J^{(1)}) 
e^{i \int d^{D} x ~ {\cal L}_1^{'}[J^{(1)},P^{(1)}, J^{(1)[2,\mu\nu]},\Phi] },
\eqa
where
\bqa
{\cal L}_1^{'} 
& = & N^2 \Bigl\{   \left( J^{(1) n } - {\cal J}^{(0)m } f_m^{~~n}(0,1) \right) P_{n}^{(1)}
 - J^{(1) m } O^{(1)}_{m } + 
V[ f_m^{~~n}(0,1) P_{n }^{(1)}; {\cal J}^{(0); \{ m_i \}, \{ \nu^i_j \} }] \Bigr\}. \nn
\label{L1p}
\eqa
Here $f_m^{~~n}(0,1) \equiv f_m^{~~n}(G^{(0)},G^{(1)})$
and $O^{(1)}_{m}$ refers to covariant operators 
written in the background metric $G^{(1)\mu\nu} = J^{(1)[2,\mu\nu]}$
which is also the source for the kinetic term with two derivatives.
In the following, we will use $G^{\mu\nu}$ and $J^{[2,\mu\nu]}$ interchangeably.
The determinant becomes
\bqa
\Delta(J) & \equiv & \Delta(J,J^{[2,\mu\nu]}) \nn
& = & 
\left|
\int D \delta G ~~ \delta \left( 
 J^{n} f_n^{~[2,\mu \nu]}( G, G + \delta G) 
- (G+\delta G)^{\mu \nu}
\right)
\right|^{-1}_{G^{\mu\nu}=J^{[2,\mu\nu]}} \nn
& = & 
\left|
\int D \delta G ~~ \delta \left( 
- \int dy J^{n}(x) \frac{ \delta f_n^{~[2,\mu \nu]}(x)}{ \delta G^{\alpha \beta}(y) } 
\delta G^{\alpha \beta}(y) - \delta G^{\mu \nu}(x) 
\right)
\right|^{-1}_{G^{\mu\nu}=J^{[2,\mu\nu]}}  \nn
& = & 
\det \left[
\delta_{(\alpha \beta)}^{(\mu \nu)} \delta(x-y)
+ J^{n}(x) \frac{ \delta f_n^{~[2,\mu \nu]}(x) }{ \delta G^{\alpha \beta}(y) } 
\right]_{G^{\mu\nu}=J^{[2,\mu\nu]}},
\label{det}
\eqa
where $ \frac{ \delta f_m^{~n}(x)}{\delta G^{\mu \nu}(y) } 
= \left. \frac{ \delta f_m^{~n}(G,G^{'}) }{\delta G^{\mu \nu}(y) } \right|_{G^{'}=G}$
and we used the fact that
$ \left. \frac{  \delta f_m^{~n}(G,G^{'}) }{ \delta G^{\mu \nu}(y) } \right|_{G^{'}=G}
= -  \left. \frac{  \delta f_m^{~n}(G,G^{'}) }{ \delta G^{'\mu \nu}(y) } \right|_{G^{'}=G}$.
$\delta_{(\alpha \beta)}^{(\mu \nu)}$ is the Kronecker delta function for symmetrized indices,
and $\delta(x-y)$ is the $D$-dimensional Dirac delta function.
To obtain the expression in the third line from the second line in Eq. (\ref{det}),
we use the fact that there is one and only one solution for the gauge fixing condition (see Appendix A).

\section{Coarse Graining}
Now we perform a coarse graining 
by integrating out high energy modes of the matrix field $\Phi$.
Although the sources $J^{(1)m}$ are also dynamical fields, one can treat them as background fields when one integrates out high energy modes of $\Phi$. 
We focus on the functional integration of the original dynamical field $\Phi$ which is coupled to the sources $J^{(1)m}$,
\bqa
Z_\Phi[J^{(1)}] & \equiv & \int D \Phi ~ 
e^{i \int d^{D} x \left[ N \tr (\Phi {\cal M}_{J^{(1)}} \Phi) 
+ U_{J^{(1)}}[\Phi] \right] },
\eqa
where ${\cal M}_{J^{(1)}}$ is the kernel for the quadratic action 
that includes the two and higher derivative terms,
\bqa
{\cal M}_{J^{(1)}} 
& = & 
-  \sqrt{ |G^{(1)}|} \Bigl[
J^{(1)[2]} + 
G^{(1)\mu \nu}
\nabla_{\mu}\nabla_{\nu} 
+
\sum_{n=3}^\infty 
 J^{(1) [2,\mu_1...\mu_n]} 
\nabla_{\mu_1}...\nabla_{\mu_n} 
\Bigr]
\label{kernel}
\eqa
and $U_{J^{(1)}}[\Phi]$ includes all other single-trace operators,
which are at least cubic in $\Phi$.
There is no operator linear in $\Phi$ because $\Phi$ is traceless.
The sources for the higher derivative terms in Eq. (\ref{kernel}) 
have the engineering scaling dimension
$[  J^{[2,\mu_1...\mu_n]}  ] = -(n-2)$,
and the mass scales associated with the sources play the role of UV cut-offs.
It is interesting to note that there are in general many scales.
Moreover, the cut-off scales are fluctuating because the sources are dynamical.
Each configuration of $ J^{[2,\mu_1...\mu_n]} $ describes a theory of $\Phi$ 
with a different set of UV cut-off scales.
We perform a real space RG transformation\cite{POLCHINSKI84,POLONYI} 
by lowering {\it some} of these energy scales.
For this, an auxiliary traceless real symmetric matrix field $\td \Phi$ is added to the original theory,
\bqa
Z_\Phi[J^{(1)}] & = & [ \det \td {\cal M} ]^{\frac{(N+2)(N-1)}{4}}  
\int D \Phi D \td \Phi ~ 
e^{i \int d^{D} x \left[ 
N \tr (\Phi {\cal M}_{J^{(1)}} \Phi) 
+ N \tr (\td \Phi \td {\cal M} \td \Phi) 
+ U_{J^{(1)}}[\Phi] \right] },
\eqa
where $\td {\cal M}$ is an arbitrary kernel for the auxiliary field.
Here $\frac{(N+2)(N-1)}{2}$ is the number of independent
components of a real traceless symmetric matrix.
We go into a new basis $\phi$ and $\tilde \phi$, 
\bqa
\Phi({\bf x}) &=& \phi({\bf x}) + \tilde \phi({\bf x}), \nn
\tilde \Phi({\bf x}) &=& \int d {\bf y} ~ \left( A({\bf x},{\bf y}) \phi({\bf y}) + B({\bf x},{\bf y}) \tilde \phi({\bf y}) \right),
\label{NB}
\eqa
where the functions $A$ and $B$ are uniquely chosen from the conditions
that the low energy field $\phi$ and the high energy field $\td \phi$
do not mix at the quadratic level,
and that the low energy field has a set of  
UV-cut off scales which are smaller than 
those for the original field $\Phi$.
Then the partition function takes the form of
\bqa
Z_\Phi[J^{(1)}] & = & 
[\det  {\cal S} \det {\cal M}_{J^{(1)'}} \det {\cal M}_{J^{(1)}}^{-1}  ]^{\frac{(N+2)(N-1)}{4}}
\int D \phi D \td \phi ~ 
e^{i \int d^{D} x \left[ 
N \tr (\phi {\cal M}_{J^{(1)'}} \phi) 
+ N \tr (\td \phi  {\cal S} \td \phi) 
+ U_{J^{(1)}}[\phi+\td \phi] \right] }, \nn
\eqa
where the quadratic action for the low energy mode is given by
\bqa
{\cal M}_{J^{(1)'}} 
& = & 
-  \sqrt{ |G^{(1)}|} \Bigl[
J^{(1)[2]} + 
G^{(1)\mu \nu}
\nabla_{\mu}\nabla_{\nu} 
+
\sum_{n=3}^\infty 
 J^{(1)' [2,\mu_1...\mu_n]} 
\nabla_{\mu_1}...\nabla_{\mu_n} 
\Bigr].
\eqa
Here the rescaled sources are
\bqa
 J^{(1)' [2,\mu_1...\mu_n]}(x)  & = & 
 e^{ c_n \alpha^{(1)} (x) dz}  J^{(1) [2,\mu_1...\mu_n]}(x),
\label{scaling}
\eqa
where $c_n$ is a set of constants which determine
how we rescale the set of UV cut-off scales.
Changing the sources for high derivative terms of $\phi$ in this way
is equivalent to lowering the UV cut-off scale 
associated with the source $J^{(1)[2,\mu_1,..,\mu_n]}$
by a factor of $e^{-c_n \alpha^{(1)} dz}$,
where $\alpha^{(1)}$ is the rate at which the UV cut-off is lowered,
and $dz$ is an infinitesimal constant.
It is noted that one can choose different speeds of coarse graining at different points in spacetime, 
and $\alpha^{(1)}(x)$ is in general position dependent.
This is a local RG procedure where
the speed of coarse graining is spacetime dependent.
Specifying $\{ c_n \}$ corresponds to choosing a particular RG scheme. 
One natural choice would be $c_n = (n-2)$ 
which reflects the fact
that $J^{(1) [2,\mu_1...\mu_n]}$ has dimension $-(n-2)$.
This amounts to lowering all UV cut-off scales in the same way.
However, this choice is not ideal for our purpose : 
the ratio between the determinants of the 
original field and the low energy field,
$[ \det {\cal M}_{J^{(1)'}} \det {\cal M}_{J^{(1)}}^{-1}] $ 
is divergent.
To illustrate this, let us consider 
the simple case where $J^{(1) [2,\mu_1...\mu_n]}$ 
and $\alpha^{(1)}(x)$ are independent of $x$.
In this case, the ratio between the determinants is given by
\bqa
\ln [ \det {\cal M}_{J^{(1)'}} \det {\cal M}_{J^{(1)}}^{-1} ]
& \sim & \alpha^{(1)} dz \int d^Dk ~~
\frac{\sum_n c_n i^n J^{(1) [2,\mu_1...\mu_n]} k_{\mu_1} ... k_{\mu_n}  }
 {\sum_n i^n J^{(1) [2,\mu_1...\mu_n]} k_{\mu_1} ... k_{\mu_n}  },
\label{d1}
\eqa
which is divergent if $c_n = (n-2)$.
In the conventional RG procedure,
the determinants do not play an important role,
and one can ignore the divergent determinant
in computing beta functions.
In our case, the determinants are important because
they provides a non-trivial action 
for the dynamical sources.
This difference comes from the fact that sources are dynamical fields
in our approach, instead of constants.
In order to avoid the UV divergence,
we choose, among many other choices, the following prescription,
\bqa
c_n & = & (n-2) ~~~\mbox{for $n \leq n_c$}, \nn
    & = & 0 ~~~\mbox{for $n > n_c$},
\label{rs}
\eqa 
where $n_c$ is a large but fixed number.
For sufficiently large momenta, Eq. (\ref{d1}) becomes
\bqa
\ln [ \det {\cal M}_{J^{(1)'}} \det {\cal M}_{J^{(1)}}^{-1} ]
& \sim & \alpha^{(1)} dz \int d^Dk ~~
\frac{c_{n_c} J^{(1) [2,\mu_1...\mu_{n_c}]} k_{\mu_1} ... k_{\mu_{n_c}}  }
 {\sum_n i^n J^{(1) [2,\mu_1...\mu_n]} k_{\mu_1} ... k_{\mu_n}  },
\label{d2}
\eqa
which is finite.
This is a well defined 
coarse graining procedure,
where we are rescaling the sources for 
higher derivative terms upto the $n_c$-th order 
to eliminate high energy mode
while avoiding the divergence in the ratio of the determinants.
In a sense, we are performing a coarse graining with two sets of scales.
The first set of scales associated with the high derivative terms up to the $n_c$-th order
plays the role of the usual UV cut-off that is rescaled to thin out high energy modes.
The second set of scales associated with the high derivative terms with more than $n_c$ derivatives
cuts off the UV divergence in the ratio of the determinants.
The conventional scheme is reproduced when $n_c$ is taken to be infinite.
It is emphasized that the specific form of rescaling in Eq. (\ref{rs})
is not important.
There exist many other schemes that regularize the divergences in the determinants.
What follows below is independent of the specific choice.
The propagator of the high energy mode is given by the difference 
between the propagators of the original field and the low energy field
\bqa
{\cal S}^{-1} & = & {\cal M}_{J^{(1)}}^{-1} - {\cal M}_{J^{(1)'}}^{-1}.
\eqa
Therefore the propagator of the high energy mode is $O(dz)$.

\begin{figure}[h!]
\centering
      \includegraphics[height=4cm,width=12cm]{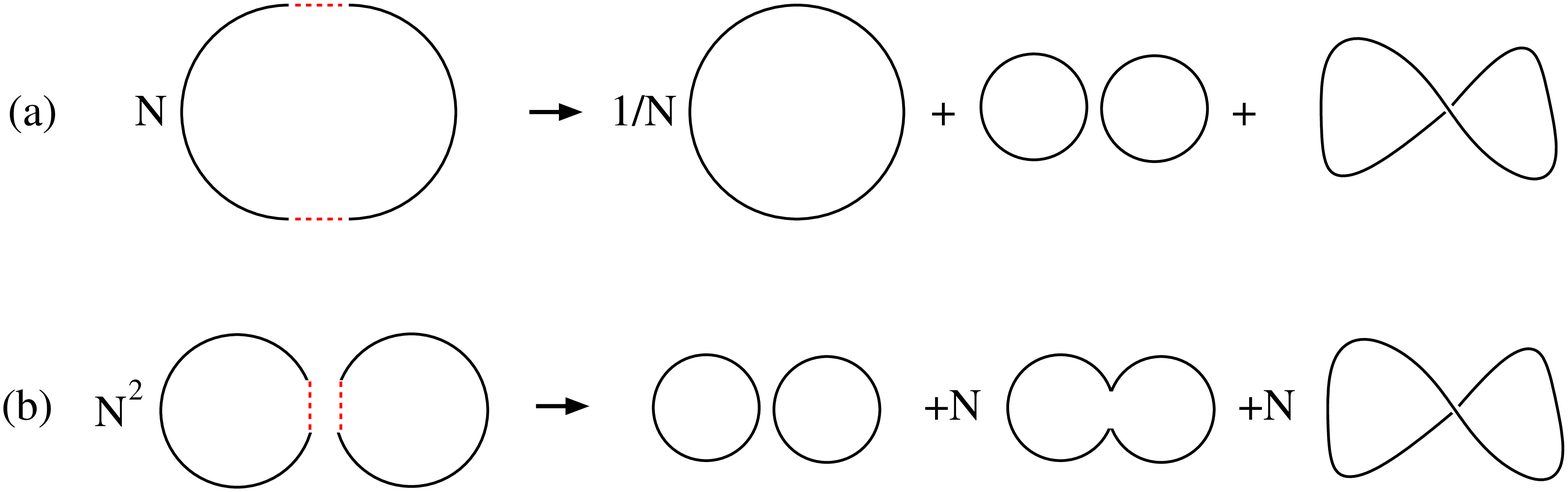}
\caption{
Two ways of generating quantum corrections to the linear order in $dz$.
Each circle denotes trace of a chain of matrix fields.
Solid lines represent chains of low energy fields
and each dashed line represents a high energy field.
(a) Contraction of a pair of high energy fields within a single-trace operator
generates two singe-trace operators (the first and the third) and one double-trace operator (the second). 
In the large $N$ limit, only the second term is $O(N^2)$.
(b) At the quadratic order, one can fuse two single-trace operators each of which contains one high energy mode.
This leads to one double-trace operator and two single-trace operators, 
all of which are $O(N^2)$.
}
\label{fig:1}
\end{figure}

Integrating out the high energy mode, we obtain 
an effective theory for the low energy mode,
\bqa
Z_\Phi[J^{(1)}] & = & 
[\det {\cal M}_{J^{(1)'}} \det {\cal M}_{J^{(1)}}^{-1}  ]^{\frac{(N+2)(N-1)}{4}}
\int D \phi ~ 
e^{i \int d^{D} x \left[ 
N \tr (\phi {\cal M}_{J^{(1)'}+\delta J^{(1)}} \phi) 
+ U_{J^{(1)}+\delta J^{(1)}}[\phi] 
+ \delta W^{(1)}[\phi]
\right] }, \nn
\eqa
where $\delta J^{(1)}$ is the quantum correction to the sources for the single-trace operators, 
and $\delta W^{(1)}$ denotes double-trace operators generated from quantum corrections.
Because the propagator of the high energy mode $\td \phi$ is of the order of $dz$,
only two diagrams contribute to the quantum corrections to the linear order in $dz$.
The first contribution comes from contracting two high energy fields within one single-trace operator
as is shown in Fig. \ref{fig:1} (a),
\bqa
\delta {\cal L} & = & -N \sqrt{ |G^{(1)}| } J^{(1) [q+1, \{ \mu^i_j \}  ]}   
\Bigl<
\tr \Bigl[
\phi
\left( \nabla_{\{\mu^1\}} \phi \right)
..
\left( \nabla_{\{\mu^{i-1}\}} \phi \right)
\left( \nabla_{\{\mu^i\}} \td \phi \right)
\left( \nabla_{\{\mu^{i+1}\}} \phi \right)
..    \nn
&& ..
\left( \nabla_{\{\mu^{j-1}\}} \phi \right)
\left( \nabla_{\{\mu^j\}} \td \phi \right)
\left( \nabla_{\{\mu^{j+1}\}} \phi \right)
..
\left( \nabla_{\{\mu^{q}\}} \phi \right)
\Bigr]
\Bigr>_{\td \phi},
\eqa
where 
$\left( \nabla_{\{\mu^i\}} \phi \right)
\equiv \left( \nabla_{\mu^i_1}^{G^{(1)}} \nabla_{\mu^i_2}^{G^{(1)}}.. \nabla_{\mu^i_{p_i}}^{G^{(1)}} \phi \right)$
and 
\bqa
< O >_{\td \phi} \equiv
\frac{ 
\int D \td \phi ~ O ~
e^{i N \int d^{D} x  \tr (\td \phi  {\cal S} \td \phi)  }
}
{ 
\int D \td \phi ~
e^{i N \int d^{D} x  \tr (\td \phi  {\cal S} \td \phi)  }
}. 
\eqa
Using the expression for the propagator of the high energy mode,
\bqa
< \td \phi_{ab}(x) \td \phi_{cd}(y) >_{\td \phi} 
& = & 
i \frac{  {\cal S}^{-1}(x,y) + {\cal S}^{-1}(y,x)  }{8N}
\left[
( \delta_{ac} \delta_{bd}
+\delta_{ad} \delta_{bc} )
-\frac{2}{N}\delta_{ab} \delta_{cd}
\right],
\eqa
one obtains
\bqa
\delta {\cal L} & = & - i \frac{\sqrt{ |G^{(1)}| } }{8} 
J^{(1) [q+1, \{ \mu^i_j \}  ]}   
\nabla_{\{\mu^i\}}^x 
\nabla_{\{\mu^j\}}^y 
[ {\cal S}^{-1}(x,y) + {\cal S}^{-1}(y,x) ]_{y=x} \nn
&&
\tr \Bigl[
\phi
\left( \nabla_{\{\mu^1\}} \phi \right)
..
\left( \nabla_{\{\mu^{i-1}\}} \phi \right)
\left( \nabla_{\{\mu^{j+1}\}} \phi \right)
..   
\left( \nabla_{\{\mu^{q}\}} \phi \right) 
\Bigr] 
\tr \Bigl[
\left( \nabla_{\{\mu^{i+1}\}} \phi \right)
..
\left( \nabla_{\{\mu^{j-1}\}} \phi \right)
\Bigr]. 
\label{c1}
\eqa
Here we drop the contributions that are sub-leading in $1/N$.
Integrating by part, if necessary, one obtains 
operators of the form,
\bqa
&& N^2 \frac{ \delta_{\alpha^{(1)}} J^{(1) m n \{ \mu \} \{ \nu \} } }{ \sqrt{ |G^{(1)}| }}
( \nabla_{ \{ \mu \} }^{(1)} O^{(1)}_{m } ) 
( \nabla_{ \{ \nu \} }^{(1)} O^{(1)}_{n } ) \nn
&& \equiv 
N^2 \sum_{p,q=0}^\infty  
\frac{ 
\delta_{\alpha^{(1)}} J^{(1) m n; \mu_1, .., \mu_p; \nu_1, .., \nu_q} 
 }{ \sqrt{ |G^{(1)}| }}
( \nabla_{ \mu_1 }^{(1)} ..\nabla_{ \mu_p }^{(1)}  O^{(1)}_{m } )
( \nabla_{ \nu_1 }^{(1)} ..\nabla_{ \nu_q }^{(1)}  O^{(1)}_{n } ).
\label{dt}
\eqa
This expression represents double-trace operators 
when both $O_m^{(1)}$ and $O_n^{(1)}$ are non-trivial.
In the special case where $O_m^{(1)} = 1$ with $p=0$
(or $O_n^{(1)} = 1$ with $q=0$), 
it reduces to single-trace operators.
The special case occurs when two high energy fields 
are adjacent to each other in a chain of matrix fields
within a single-trace operator, 
as in $J^{[4,\mu \nu]} \sqrt{ |G^{(1)}| } \frac{1}{N} \tr[ \td \phi \td \phi \phi \nabla_\mu \nabla_\nu \phi]$.
The second contribution comes from fusing two single-trace operators
as is shown in Fig. \ref{fig:1} (b),
\bqa
\delta {\cal L}^{'} & = & 
i \frac{N^2}{2} 
\int d^D y ~
\sqrt{ |G^{(1)}(x)| } 
\sqrt{ |G^{(1)}(y)| } 
J^{(1) [q+1, \{ \mu^i_j \}  ]}(x)
J^{(1) [p+1, \{ \nu^i_j \}  ]}(y) \nn
&&
\Bigl<
\tr \Bigl[
\phi
\left( \nabla_{\{\mu^1\}} \phi \right)
..
\left( \nabla_{\{\mu^{i-1}\}} \phi \right)
\left( \nabla_{\{\mu^i\}} \td \phi \right)
\left( \nabla_{\{\mu^{i+1}\}} \phi \right)
..   
\left( \nabla_{\{\mu^{q}\}} \phi \right)
\Bigr]_x \nn
&& \tr \Bigl[
\phi
\left( \nabla_{\{\nu^1\}} \phi \right)
..
\left( \nabla_{\{\nu^{j-1}\}} \phi \right)
\left( \nabla_{\{\nu^j\}} \td \phi \right)
\left( \nabla_{\{\nu^{j+1}\}} \phi \right)
..   
\left( \nabla_{\{\nu^{p}\}} \phi \right)
\Bigr]_y
\Bigr>_{\td \phi}.
\eqa
Contracting the high energy modes,
one obtains both single-trace and double-trace operators,
\bqa
&& \delta {\cal L}^{'}  =  
-\frac{1}{16} 
\int d^D y ~
\sqrt{ |G^{(1)}(x)| } 
\sqrt{ |G^{(1)}(y)| } 
J^{(1) [q+1, \{ \mu^i_j \}  ]}(x)
J^{(1) [p+1, \{ \nu^i_j \}  ]}(y) \times \nn
&&
\nabla_{\{\mu^i\}}^x 
\nabla_{\{\nu^j\}}^y 
[ {\cal S}^{-1}(x,y) + {\cal S}^{-1}(y,x) ] 
\Bigl\{
-2 \tr \Bigl[
\phi
\left( \nabla_{\{\mu^1\}} \phi \right)
..
\left( \nabla_{\{\mu^{i-1}\}} \phi \right)
\left( \nabla_{\{\mu^{i+1}\}} \phi \right)
..   
\left( \nabla_{\{\mu^{q}\}} \phi \right) \Bigr] \times \nn
&&  ~~~~~~~~~~ \tr \Bigl[
\phi^{'}
\left( \nabla_{\{\nu^1\}} \phi^{'} \right)
..
\left( \nabla_{\{\nu^{j-1}\}} \phi^{'} \right)
\left( \nabla_{\{\nu^{j+1}\}} \phi^{'} \right)
..   
\left( \nabla_{\{\nu^{p}\}} \phi^{'} \right)
\Bigr] \nn
&& + N 
 \tr \Bigl[
\phi
\left( \nabla_{\{\mu^1\}} \phi \right)
..
\left( \nabla_{\{\mu^{i-1}\}} \phi \right)
\left( \nabla_{\{\nu^{j-1}\}} \phi^{'} \right)
\left( \nabla_{\{\nu^{j-2}\}} \phi^{'} \right)
..
\left( \nabla_{\{\nu^{j+1}\}} \phi^{'} \right)
\left( \nabla_{\{\mu^{i+1}\}} \phi \right)
..   
\left( \nabla_{\{\mu^{q}\}} \phi \right) \Bigr] \nn
&&
+ N
\tr \Bigl[
\phi
\left( \nabla_{\{\mu^1\}} \phi \right)
..
\left( \nabla_{\{\mu^{i-1}\}} \phi \right)
\left( \nabla_{\{\nu^{j+1}\}} \phi^{'} \right)
\left( \nabla_{\{\nu^{j+2}\}} \phi^{'} \right)
..
\left( \nabla_{\{\nu^{j-1}\}} \phi^{'} \right)
\left( \nabla_{\{\mu^{i+1}\}} \phi \right)
..   
\left( \nabla_{\{\mu^{q}\}} \phi \right) 
\Bigr] 
\Bigr\}. \nn
\label{c2}
\eqa
In this expression, it is understood that $\phi = \phi(x)$ and $\phi^{'}=\phi(y)$.
Although the quantum corrections appear to be non-local,
the propagator for the high energy mode decays exponentially in real space, 
allowing one to do a gradient expansion.
The scale that controls the gradient expansion is the UV cut-off
which is set by the dynamical sources.
This results in local double-trace operators of the form in Eq. (\ref{dt})
and quantum corrections to single-trace operators.
The determinant 
$[\det {\cal M}_{J^{(1)'}} \det {\cal M}_{J^{(1)}}^{-1}  ]^{\frac{(N+2)(N-1)}{4}}$ 
gives rise to a Casimir energy 
that depends on the source $J^{(1)}$.
The Casimir energy provides a `potential' energy 
for the dynamical sources while
the double-trace operators becomes
a quadratic `kinetic' term for the conjugate fields
as will be shown in Sec. VI and VII.

Note that even though the action for $\Phi$ in Eq. (\ref{L1p}) has only single-trace operators,
double-trace operators are generated 
in the renormalized action\cite{HEEMSKERK10,Faulkner1010}. 
Triple or higher trace operators are at least order of $dz^2$
and can be ignored in the small $dz$ limit.
After exponentiating the determinant into a quantum action for the dynamical sources, 
the total Lagrangian for the low energy field and dynamical sources can be written in the following form,
\bqa
{\cal L}_2 
& = & 
N^2 \Bigl\{ V[ f_m^{~~n}(0,1) P_{n }^{(1)}; {\cal J}^{(0); \{ m_i \}, \{ \nu^i_j \} }] \nn
&& +  \left( J^{(1) n } - {\cal J}^{(0)m } f_m^{~~n}(0,1) \right) P_{n}^{(1)} + \delta_{\alpha^{(1)}} {\cal L}[ J^{(1) m } ] \nn
&& - ( J^{(1) m } + \delta_{\alpha^{(1)}} J^{(1) m \{ \mu \}} \nabla_{ \{ \mu \} }^{(1)} ) O^{(1)}_{m } 
+ \frac{ \delta_{\alpha^{(1)}} J^{(1) m n \{ \mu \} \{ \nu \} } }{ \sqrt{ |G^{(1)}| }}
 ( \nabla_{ \{ \mu \} }^{(1)} O^{(1)}_{m } ) 
( \nabla_{ \{ \nu \} }^{(1)} O^{(1)}_{n } )  \Bigr\}. 
\label{45}
\eqa
Here 
$\int d^{D} x ~ \delta_{\alpha^{(1)}} {\cal L}[ J^{(1) m } ]  
= -i\frac{(N+2)(N-1)}{4N^2} \ln [\det {\cal M}_{J^{(1)'}} \det {\cal M}_{J^{(1)}}^{-1}  ]$.
We use the same notation $O_m^{(1)}$ 
to represent single-trace operators constructed from the low energy field $\phi$.
$ \delta_{\alpha^{(1)}} J^{(1) m \{ \mu \} }$ includes both the change
in the quadratic sources caused by lowering UV cut-off 
in Eq. (\ref{scaling})
and the quantum corrections to the single-trace sources
in Eqs. (\ref{c1}) and (\ref{c2}).
$\delta_{\alpha^{(1)}} J^{(1) m n \{ \mu \} \{ \nu \} } $ denotes the sources for double-trace 
operators generated from quantum corrections
in Eqs. (\ref{c1}) and (\ref{c2}).
In  the Casimir energy and the quantum corrections to the sources, 
it is enough to keep only those contributions to the order of $dz$. 
Therefore, $\delta_{\alpha^{(1)}} {\cal L}[ J^{(1) m } ]$,
$\delta_{\alpha^{(1)}} J^{(1) m \{ \mu \} }$ and
$ \delta_{\alpha^{(1)}} J^{(1) m n \{ \mu \} \{ \nu \} }$
are linear in $\alpha^{(1)} dz$.
In general, all terms that respect 
the $D$-dimensional diffeomorphism invariance are allowed in 
the Casimir energy and the quantum corrections to the sources, 
\bqa
 \delta_{\alpha^{(1)}} {\cal L}[ J^{(1) m } ] & = & dz ~ \alpha^{(1)}(x) \sqrt{|G^{(1)}|}
\left\{
C_0[ J^{(1)} ] + C_1[ J^{(1)} ] {\cal R} + ...
\right\}, \label{casimiar} \\
\delta_{\alpha^{(1)}} J^{(1) m \{ \mu \} } & = & dz ~ \alpha^{(1)}(x)
A^{m \{ \mu \} }[ J^{(1)} ], \label{qc1} \\
\delta_{\alpha^{(1)}} J^{(1) m n \{ \mu \} \{ \nu \} } 
& = & dz ~ \alpha^{(1)}(x)
B^{ m n \{ \mu \} \{ \nu \}}[ J^{(1)} ]. \label{qc2} 
\eqa
Here $ \delta_{\alpha^{(1)}} {\cal L}[ J^{(1) m } ] $ is the Casimir energy 
which includes the cosmological constant $C_0$ and 
the $D$-dimensional Ricci scalar ${\cal R}$.
It also includes higher order terms in the curvature tensor
and derivative action for other sources,
$J^{[2,\mu_1,..,\mu_n]}$.
Since the Casimir energy comes from the determinant 
of the quadratic operator of $\Phi$, 
it depends only on the sources for the quadratic operators.
On the other hand, 
in Eqs. (\ref{qc1}) and (\ref{qc2}),
cubic or higher order operators 
are also renormalized by quantum corrections.
$C_0$, $C_1$, $A^{m \{ \mu \} }$ and $B^{mn \{ \mu \} \{ \nu \} }$ 
are finite functions of the sources $J^{(1)}$ 
and their covariant derivatives.
Note that one needs to include descendant operators 
for quantum corrections in Eq. (\ref{45})
to express 
$\delta_{\alpha^{(1)}} J^{(1) m \{ \mu \} }$ and
$ \delta_{\alpha^{(1)}} J^{(1) m n \{ \mu \} \{ \nu \} }$
as linear functions of $\alpha^{(1)}$ without derivative
as in Eqs. (\ref{qc1}) and (\ref{qc2}).

It is noted that the partition function is independent of $\alpha^{(1)}(x)$\cite{SLEE10}.
This is because $\alpha^{(1)}(x)$ is an arbitrary function introduced to 
change the  length scale in the coarse graining procedure.
One can choose any speed of RG without affecting the partition function.
If one modifies the parameter $\alpha^{(1)}$ 
to $\alpha^{(1)}+\delta \alpha$, 
the quantum corrections will be modified accordingly,
exactly undoing the changes caused by $\delta \alpha$.
In other words, one has to add quantum corrections (counter terms) 
so that the partition function computed from the low energy effective theory with a lower UV cut-off is equal to the one computed from the bare theory with the original cut-off.
This is nothing but the cut-off independence of 
the partition function in the Wilsonian RG.
The fact that the partition function is independent of 
the choice of $\alpha^{(1)}(x)$
will become important later
to obtain the $(D+1)$-dimensional diffeomorphism 
invariance in the holographic description
as will be discussed in Sec. VII.

\section{Shift}

\begin{figure}[h!]
\centering
      \includegraphics[height=4.7cm,width=6.3cm]{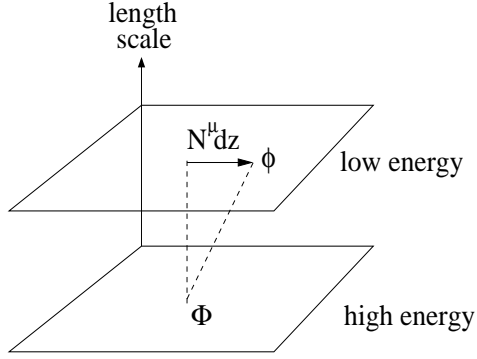}
\caption{
The $D$-dimensional coordinate of the low energy mode
is shifted infinitesimally relative to the coordinate
of the field defined at high energy.
}
\label{fig:shift}
\end{figure}

One key difference of the present construction from the conventional RG procedure\cite{POLCHINSKI84,POLONYI} is that 
the sources (and the conjugate fields) that are coupled with the low energy field  are dynamical.
In particular, the low energy field is covariantly coupled with the fully dynamical $D$-dimensional metric.
Therefore, there is a freedom to choose the coordinate system for the low energy mode without modifying the form of the Lagrangian.
To take advantage of this extra freedom, 
we will choose a new coordinate system for the low energy mode
which is infinitesimally shifted along a $D$-dimensional direction
compared to the original spacetime\cite{Douglas:2010rc}.
The point is that 
one does not have to use the same $D$-dimensional coordinate
for the low energy field as the 
one for the high energy field
as is illustrated in Fig. \ref{fig:shift}.
For this, we single out the action for the low energy mode,
\bqa
S_l & = & N^2 \int d^{D} x ~ \left[
-( J^{(1) m } + \delta_{\alpha^{(1)}} J^{(1) m \{ \mu \}} \nabla_{ \{ \mu \} }^{(1)} ) O^{(1)}_{m } 
+ \frac{ \delta_{\alpha^{(1)}} J^{(1) m n \{ \mu \} \{ \nu \} } }{  \sqrt{ |G^{(1)}| }     }
 ( \nabla_{ \{ \mu \} }^{(1)} O^{(1)}_{m } ) 
( \nabla_{ \{ \nu \} }^{(1)} O^{(1)}_{n } ) 
\right],  \nn
\eqa
and change the variable,
\bqa
\td \phi(y) &=& \phi(x)
\eqa
with $y^\mu = x^\mu  + N^{(1)\mu}(x) dz$.
In the new coordinate system, the metric and the sources
are transformed as tensors. 
The operators transform as tensor densities of weight one.

To the linear order of $dz$, we have
\bqa
S_l & = & N^2 \int d^D y ~ \left[
- \td J^{(1) m } \td O^{(1)}_{m }
- \delta_{\alpha^{(1)}} J^{(1) m \{ \mu \} } \nabla_{ \{ \mu \} }^{(1)}  O^{(1)}_{m } 
+ \frac{ \delta_{\alpha^{(1)}} J^{(1) m n \{ \mu \} \{ \nu \} } }{  \sqrt{ | G^{(1)}| }     }
( \nabla_{ \{ \mu \} }^{(1)} O^{(1)}_{m } ) 
(  \nabla_{ \{ \nu \}}^{ (1)}   O^{(1)}_{n } )  
\right], \nn
\eqa 
where  
$\td O^{(1)}_m$'s are the covariant operators constructed with the covariant derivative
$\td \nabla^{(1)}$ with the new metric
\bqa
\td G^{(1)}_{\mu \nu}(y) = 
\frac{ \partial x^\lambda}{\partial y^\mu}
\frac{ \partial x^\sigma}{\partial y^\nu}
G^{(1)}_{\lambda \sigma}(x).
\eqa
These operators are coupled with the new sources given by
\bqa
\td J^{(1) [q,\{ \mu^i_j  \}] }(y) & = & 
\left[ \prod_{i,j} \frac{ \partial y^{\mu^i_j} }{\partial x^{\nu^i_j}} \right] 
J^{(1) [q,\{ \nu^i_j  \}]}(x).
\eqa
We can ignore the shift in $\delta_{\alpha^{(1)}} J^{(1) m \{ \mu \} }$
and $\delta_{\alpha^{(1)}} J^{(1) m n \{ \mu \} \{ \nu \} }$ 
because they are already order of $dz$.
The Lagrangian in the shifted coordinate becomes
\bqa
{\cal L}_l & = & 
N^2 \Bigl\{
- \td J^{(1) m } f_m^{~n}(G^{(1)}+ \delta_{N^{(1)}} G^{(1)}, G^{(1)}) O^{(1)}_{n} 
- \delta_{\alpha^{(1)}} J^{(1) m \{ \mu \} } \nabla_{ \{ \mu \} }^{(1)} O^{(1)}_{m} 
\nn
&& + \frac{ \delta_{\alpha^{(1)}} J^{(1) m n \{ \mu \} \{ \nu \} } }{  \sqrt{ |G^{(1)}| }     } 
( \nabla_{ \{ \mu \} }^{(1)} O^{(1)}_{m } ) 
(  \nabla_{ \{ \nu \}}^{ (1)}   O^{(1)}_{n } )  \Bigr\} \nn
& = & 
N^2 \Bigl\{ - \left( J^{(1) m } + \delta_{\alpha^{(1)}} J^{(1) m \{ \mu \} } \nabla_{ \{ \mu \} }^{(1)} + \delta_{N^{(1)}} J^{(1) m }   \right) O^{(1)}_{m}  \nn
&& + \frac{ \delta_{\alpha^{(1)}} J^{(1) m n \{ \mu \} \{ \nu \} } }{  \sqrt{ |G^{(1)}| }     } 
( \nabla_{ \{ \mu \} }^{(1)} O^{(1)}_{m } ) 
(  \nabla_{ \{ \nu \}}^{ (1)}   O^{(1)}_{n } )     \Bigr\},
\eqa
where 
$\delta_{N^{(1)}} G^{(1)\mu\nu} = \td G^{(1)\mu\nu} - G^{(1)\mu\nu}$
and
\bqa
\delta_{N^{(1)}} J^{(1) m } & = & 
( \td J^{(1) m } - J^{(1) m } ) 
 + J^{(1),n} 
\Bigl( f_n^{~m}(G^{(1)}+ \delta_{N^{(1)}} G^{(1)}, G^{(1)}) - \delta_n^{~m} \Bigr).
\label{C}
\eqa
As was the case for $\alpha^{(1)}$,
the partition function is clearly independent of $N^{(1)\mu}$,
because different choices of shift merely corresponds to
different choices of coordinate system
for the low energy mode.
This completes one cycle of the RG procedure.
We have a theory of the low energy field coupled with dynamical sources
whose fluctuations are controlled by the action 
generated from the high energy mode,

\bqa
{\cal L}_2 
& = & 
N^2 \Bigl\{ V[ f_m^{~~n}(0,1) P_{n }^{(1)}; {\cal J}^{(0); \{ m_i \}, \{ \nu^i_j \} }] \nn
&&   +\left( J^{(1) n } - {\cal J}^{(0)m } f_m^{~~n}(0,1) \right) P_{n}^{(1)} + \delta_{\alpha^{(1)}} {\cal L}[ J^{(1) m } ] \nn
&& - 
\left( J^{(1) m } + \delta_{\alpha^{(1)}} J^{(1) m \{ \mu \} } \nabla_{ \{ \mu \} }^{(1)} + \delta_{N^{(1)}} J^{(1) m } 
\right) O^{(1)}_{m } \nn
&& + \frac{ \delta_{\alpha^{(1)}} J^{(1) m n \{ \mu \} \{ \nu \} } }{  \sqrt{ |G^{(1)}| }     } 
 ( \nabla_{ \{ \mu \} }^{(1)} O^{(1)}_{m } ) 
(  \nabla_{ \{ \nu \}}^{ (1)}   O^{(1)}_{n } )  
\Bigr\}.
\eqa

\section{Construction of bulk theory}
Now we repeat the procedures in Secs. III-V.
Another set of auxiliary fields are introduced 
to remove the double-trace operators for the low energy fields 
followed by the gauge fixing to obtain
\bqa
Z & = & \int 
D J^{(1)n} D P^{(1)}_n  
D J^{(2)n} D P^{(2)}_n
D  \Phi \Delta(J^{(1)}) \Delta (J^{(2)}) ~~ 
e^{i \int d^{D} x ~ {\cal L}_3 },
\label{57}
\eqa
where we use the notation $\Phi$ for the low energy mode to 
avoid introducing a new notation for the low energy mode 
at each step of RG, 
and the Lagrangian is given by 
\bqa
{\cal L}_3 
& = & N^2 \Bigl\{
V[ f_m^{~~n}(0,1) P_{n }^{(1)}; {\cal J}^{(0); \{ m_i \}, \{ \nu^i_j \} }] \nn
&& +  \left( J^{(1) n } - {\cal J}^{(0)m } f_m^{~~n}(0,1) \right) P_{n}^{(1)} + \delta_{\alpha^{(1)}} {\cal L}[ J^{(1) m } ] \nn
&& 
+ J^{(2) n} \left( P^{(2)}_n - O^{(2)}_n \right) 
- \left( J^{(1) m } + \delta_{\alpha^{(1)}} J^{(1) m \{ \mu \} } \nabla_{ \{ \mu \} }^{(1)} + \delta_{N^{(1)}} J^{(1) m } 
\right) f_m^{~~n}(1,2) P^{(2)}_{n} \nn
&& +  \frac{ \delta_{\alpha^{(1)}} J^{(1) m n \{ \mu \} \{ \nu \} } }{  \sqrt{ |G^{(1)}| }     }  
( \nabla_{ \{ \mu \} }^{(1)} f_m^{~k}(1,2) P^{(2)}_{k } )
( \nabla_{ \{ \nu \} }^{(1)} f_n^{~k^{'}}(1,2) P^{(2)}_{k^{'} } )  
 \Bigr\}. 
\eqa
Here $O_n^{(2)}$'s represent covariant operators constructed 
with the metric $G^{(2) \mu \nu} = J^{(2)[2,\mu \nu]}$.
$\Delta(J^{(2)})$ in Eq. (\ref{57}) is the Jacobian generated from the gauge fixing.
As was done in Sec. IV, high energy modes are integrated out 
with a spacetime dependent coarse graining rate $\alpha^{(2)}(x)$ 
to generate 
the Casimir energy for $J^{(2)n}$ and
another set of quantum corrections to single-trace operators
and double-trace operators,
which are proportional to 
$\alpha^{(2)}(x) dz$.
This is followed by another infinitesimal shift along the $D$-dimensional direction
$N^{(2)\mu}(x)$ as in Sec. V.
Note that the 
$\alpha^{(2)}(x)$ and $N^{(2)\mu}(x)$
are independent of 
$\alpha^{(1)}(x)$ and $N^{(1)\mu}(x)$.
Namely, we can choose different rate of coarse graining 
and different shift at each scale.
The double trace operators are again removed by
introducing a third set of auxiliary fields $J^{(3)m}$ and $P^{(3)}_m$.

If we repeat these steps $L$ times, 
the partition function can be written as
\bqa
Z & = & \int 
\Pi_{l=1}^L \left[ D J^{(l)n} D P^{(l)}_n \Delta(J^{(l)}) \right] D \Phi ~~ 
e^{i \int d^{D} x ~ {\cal L}_4 },
\eqa
where
\bqa
{\cal L}_4 
& = & 
N^2 
V[ f_m^{~~n}(0,1) P_{n }^{(1)}; {\cal J}^{(0); \{ m_i \}, \{ \nu^i_j \} }] \nn
&& + N^2 \sum_{l=0}^L \Bigl\{
  \left( J^{(l+1) n } - J^{(l)m } f_m^{~~n}(l,l+1) \right) P_{n}^{(l+1)}  \Bigr\} \nn
&& + N^2 \sum_{l=1}^L \Bigl\{
 \delta_{\alpha^{(l)}} {\cal L}[ J^{(l) m } ] 
 - \left( \delta_{\alpha^{(l)}}  J^{(l) m \{ \mu \} } \nabla_{ \{ \mu \} }^{(l)} + \delta_{N^{(l)}} J^{(l) m }  
\right) f_m^{~~n}(l,l+1) P^{(l+1)}_{n}  \nn
&& +  \frac{ \delta_{\alpha^{(l)}} J^{(l) m n \{ \mu \} \{ \nu \} } }{  \sqrt{ |G^{(l)}| }     }  
( \nabla_{ \{ \mu \} }^{(l)} f_m^{~k}(l,l+1) P^{(l+1)}_{k } ) 
( \nabla_{ \{ \nu \} }^{(l)} f_n^{~k^{'}}(l,l+1) P^{(l+1)}_{k^{'} } ) \Bigr\} \nn
&& - N^2 J^{(L+1) n} O^{(L+1)}_n. 
\eqa
Here it is understood that $J^{(0)n} = {\cal J}^{(0)n}$.

Now we first take the limit with 
$dz \rightarrow 0$
and $L \rightarrow \infty$
with fixed $z_L = L dz$,
where $z = l dz$ becomes a continuous coordinate 
that labels the length scale in the range 
$0 \leq z \leq z_L$.
$J^{(l)n}(x)$, $P^{(l)}_n(x)$, $\alpha^{(l)}(x)$ and $N^{(l)\mu}(x)$
become $D+1$-dimensional fields
$J^{n}(x,z)$, $P^{}_n(x,z)$, $\alpha^{}(x,z)$ and $N^{\mu}(x,z)$,
respectively.
Then, we take the $z_L \rightarrow \infty$ limit,
which amounts to taking the low energy limit where one
push the RG procedure to the IR limit.
In this limit, the partition function becomes
\bqa
Z[ {\cal J} ] & = & \left. 
\int {\cal D}J(x,z) {\cal D}P(x,z) {\cal M}(J) ~~ 
e^{i \Bigl( S_{UV}[P(x,0)] + S[J(x,z),P(x,z)] + S_{IR}[J(x,\infty)] \Bigr) } \right|_{J(x,0) = {\cal J}(x)}, 
\nn
\eqa
where
\bqa
{\cal D}J(x,z) {\cal D}P(x,z) & \equiv & 
\prod_{l=1}^\infty \left[ DJ^{(l)}(x) DP^{(l)}(x) \right], \nn
{\cal M}(J)  & \equiv & \prod_{l=1}^\infty  \Delta( J^{(l)} ), \nn
S_{UV} & = & N^2 \int d^{D} x ~~ V[ P_m(x,0);  {\cal J}^{(0); \{ m_i \}, \{ \nu^i_j \} } ], \nn
S & = &  N^2 \int d^{D} x \int dz ~~ \Bigl[
 ( {\cal D}_z J^n )  P_n -\alpha(x,z) {\cal H} - N^{\mu}(x,z) {\cal H}_{\mu}
\Bigr], \nn
S_{IR} & = & -i \ln \int D \Phi ~~ e^{ -i N^2 \int d^{D} x J^{m}(x,\infty) O_m^{G(x,\infty)}}.
\eqa
Here $S_{UV}$ and $S_{IR}$ are the actions defined at 
the UV ($z=0$) and the IR ($z=\infty$) boundaries respectively.
$S$ is the bulk action.
${\cal D}_z J^n \equiv  (\partial_z J^n + J^m {\cal A}_m^{~n} ) $
is a `covariant derivative' 
for a vector $J^n$ defined in the space of operators
with the connection,
\bqa
{\cal A}_m^{~n}(x,z) &= & 
\left. \partial_z f_m^{~n}(G(x,z),G(x,z^{'})) \right|_{z^{'}=z} \nn
&=& \int dy ~ \partial_z G^{\mu \nu}(y)  \frac{ \delta f_m^{~n}(x)}{\delta G^{\mu \nu}(y) }.
\eqa 
Note that ${\cal D}_z$ is not related to the covariant derivative in the $D$-dimensional spacetime $\nabla_\mu$.
${\cal H}$ and ${\cal H}_\mu$ are given by
\bqa
{\cal H} & = & A^{m \{ \mu \} }[ J(x) ] ( \nabla_{ \{ \mu \} } P_m )
- \frac{ B^{ m n \{ \mu \} \{ \nu \}}[ J(x) ]}{  \sqrt{ |G| }     } 
( \nabla_{ \{ \mu \} } P_{m} ) 
( \nabla_{ \{ \nu \} } P_{n} )  \nn
&&
- \sqrt{|G|}
\Bigl\{
C_0[J(x)] + C_1[J(x)] {\cal R} + ...
\Bigr\}, 
\label{sh}
\\
{\cal H}_\mu & = & 
-2 \sqrt{|G|} \nabla^\nu 
\left[
\frac{1}{\sqrt{|G|}}
\left(
 P_{[2,\mu\nu]}(x) +
\int dy ~ J^n(y)  \frac{ \delta f_n^{~m}(y)}{\delta G^{\mu \nu}(x) } P_m(y)
\right)
\right] \nn
&& - \sum_{[q,\{ \mu^i_j \}] \neq [2,\mu\nu]}
\Bigl[
\sqrt{|G|}
\sum_{a,b}
\nabla_\nu \Bigl(
\frac{1}{\sqrt{|G|}}
J^{[q,
\{ \mu^1_1 \mu^1_2 ... \mu^a_{b-1} \nu \mu^a_{b+1} ... \}
]}
P_{[q,\{ \mu^1_1 \mu^1_2 ... \mu^a_{b-1} \mu \mu^a_{b+1} ... \}]} 
\Bigr) \nn
&& ~~~~~~~~~~~~~~~~~
 + ( \nabla_\mu J^{[q,\{ \mu^i_j \}]} )
 P_{[q,\{ \mu^i_j \}]} 
\Bigr]
\label{sm}
\eqa
with $\mu=0,1,..,(D-1)$.

\begin{figure}[h!]
\centering
      \includegraphics[height=6cm,width=14cm]{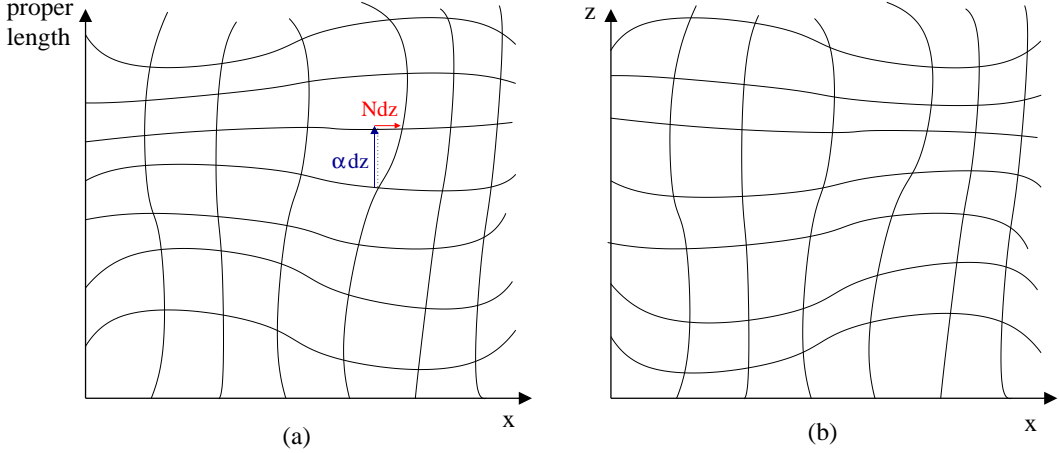}
\caption{
(a) 
Bulk spacetime made of the $D$-dimensional boundary spacetime and 
the semi-infinite line 
that represents the length scale 
in the RG procedure.
Each step of coarse graining, 
say the $l$-th step, 
generates a set of $D$-dimensional fields 
$\left( J^{(l)n}(x), P^{(l)}_n(x) \right)$
that represent dynamical sources and operators
at that scale.
These fields  are combined into $(D+1)$-dimensional fields 
$\left( J^{n}(x,z), P_n(x,z) \right)$
in the bulk, 
where the extra coordinate is given by $z = l dz$.
Each `vertical' line traces the positions of 
the bulk fields 
which are generated from the original field variable $\Phi(x)$ 
at each $x$ in the boundary spacetime. 
The spacetime dependent shift $N^{\mu}(x,z)$ 
causes the bulk fields to have different $D$-dimensional coordinates
from that of $\Phi(x)$.
Each `horizontal' line represents the manifold in the bulk spacetime with an equal $z$ coordinate.
Because the speeds of coarse graining are in general 
different at different points in spacetime, 
two points within the manifold  with an equal $z$
do not in general have the same proper length along the extra dimension,
where the proper length is the scale in the RG.
(b) 
The same bulk spacetime where the coordinate $z$ is used instead of the proper length
along the extra dimension.
The vertical lines have the same meaning as in (a).
Each horizontal line represents the manifold with an equal proper length, that is, 
the set of points with the same length scale in RG.
Note that an horizontal line that is concave upward in (a) is concave downward in (b).
}
\label{fig:2}
\end{figure}

Starting from the $D$-dimensional matrix field theory,
we obtained a $(D+1)$-dimensional theory for dynamical sources
and operators.
The extra dimension parameterized by $z$ 
represents the length scale 
in the RG.
One can choose different speed $\alpha(x,z)$ of RG 
at different points in spacetime and scale.
Therefore $z$ is not a gauge invariant quantity.
What is gauge invariant is the length scale whose
infinitesimal increment is given by  $d \tau = \alpha dz$.
In this sense, the physical length scale can be viewed as a proper length
along the direction of the extra dimension.
This is illustrated in Fig. \ref{fig:2}.
The theory in the bulk has 
the dynamical metric and its conjugate momentum 
as dynamical degrees of freedom.
Therefore it is natural to expect that
the bulk theory is a gravitational theory.
In the next section, we will see that the theory in the bulk
indeed respects the $(D+1)$-dimensional diffeomorphism
invariance.


\section{Hamiltonian gravity}
If the RG scale $z$ is interpreted as  `time',
the theory can be viewed as a Hamiltonian system
where the sources $J^m$'s play the role of `coordinates'
and the operators $P_m$'s are the `momenta'.
The sources and operators are conjugate to each other as expected.
The `Hamiltonian' is given by
\bqa
{\bf H} & = & \sum_{M=0}^{D} \int d^{D} x ~ N^M {\cal H}_M,
\label{H}
\eqa
where $N^{D}(x,z) \equiv \alpha(x,z)$ and 
${\cal H}_{D} \equiv {\cal H}$.
Note that the `time' $x^{D}=z$ is different from the real time $x^0$
in the boundary field theory.
The Hamiltonian in Eq. (\ref{H}) generates the evolution along 
the time $x^{D}$ associated with increasing length scale of the system, 
not along the real time $x^0$.
In this sense, one can regard the Hamiltonian as a generator
for a {\it quantum beta function}.
One difference from the usual Hamiltonian system
is that the `covariant' derivative ${\cal D}_z$ is used in the action.
The non-trivial connection originates from the fact 
the fields $J^n(x,z)$, $P_n(x,z)$ defined at different length scales in general have different metric.
Because the definition of the covariant operators and their sources is tied with the metric, 
a change in metric effectively induces changes in all sources.
Physically, the momentum canonically conjugate 
to the metric is the energy-momentum tensor 
given by $\pi_{[2,\mu\nu]} = \frac{1}{N^2} \frac{\delta S}{\delta G^{\mu \nu}}$.
There are many other contributions to the energy momentum tensor
besides $P_{[2,\mu\nu]} = \frac{\sqrt{|G|}}{N} \tr [ \Phi \nabla_\mu \nabla_\nu \Phi]$
because metric enters not only in  
$\frac{\sqrt{|G|}}{N} \tr [ \Phi \nabla_\mu \nabla_\nu \Phi]$
but also in the definition of all other covariant operators.
This suggests that  $P_{[2,\mu\nu]}$ is not the canonical momentum of the metric,
which is also reflected in the non-trivial measure $\Delta(J)$ 
in the functional integration,
and the unconventional form of the shift for the metric in Eq. (\ref{sm}).
In order to go to the canonical basis,
we define a new momentum
for the metric and keep the same conjugate momenta
for all other variables,
\bqa
\pi_{[2,\mu \nu]}(x) & = & P_{[2,\mu\nu]}(x) +
\int dy ~ J^n(y)  \frac{ \delta f_n^{~m}(y)}{\delta G^{\mu \nu}(x) } P_m(y), 
\label{pi1}
\\
\pi_m & = & P_m, ~~~ \mbox{for $m \neq [2,\mu\nu]$}.
\label{pi2}
\eqa
The last term in Eq. (\ref{pi1}) takes into account 
the metric dependence in general operators.
The Jacobian from the change of variable,
\bqa
\left| \frac{ \delta P_{[2,\mu\nu]}(y)}{\delta \pi_{[2,\alpha\beta]}(x)} \right| 
& = & \det
\left[
\delta_{(\alpha \beta)}^{(\mu \nu)} \delta(x-y)
+ J^{n}(y) \frac{ \delta f_n^{~[2,\mu \nu]}(y)}{ \delta G^{\alpha \beta}(x) } 
\right]^{-1} \nn
& = & \Delta(J)^{-1}
\eqa
exactly cancels $\Delta(J)$ in the measure.
The partition function and 
the action takes the canonical form\cite{ADM} 
in the new variables,
\bqa
Z[ {\cal J} ] & = & \left. 
\int {\cal D}J(x,z) {\cal D}\pi(x,z) ~~ 
e^{i \Bigl( S_{UV}[\pi(x,0)] + S[J(x,z),\pi(x,z)] + S_{IR}[J(x,\infty)] \Bigr) } \right|_{J(x,0) = {\cal J}(x)}, 
\eqa
where
\bqa
S & = &  N^2 \int d^{D} x dz ~~ \Bigl[
 ( \partial_z J^n )  \pi_n - \alpha(x,z) {\cal H} - N^{\mu}(x,z) {\cal H}_{\mu}
\Bigr]. 
\eqa
Note that $D_z$ is replaced by the usual derivative in the canonical variables. 
Moreover, the `momentum constraint' ${\cal H}_\mu$ that generates the $D$-dimensional shift 
takes the standard form,
\bqa
{\cal H}_\mu & = &
-2 \nabla^\nu \pi_{[2,\mu \nu]} 
- \sum_{[q,\{ \mu^i_j \}] \neq [2,\mu\nu]}
\Bigl[
\sum_{a,b}
\nabla_\nu \Bigl(
J^{[q,
\{ \mu^1_1 \mu^1_2 ... \mu^a_{b-1} \nu \mu^a_{b+1} ... \}
]}
\pi_{[q,\{ \mu^1_1 \mu^1_2 ... \mu^a_{b-1} \mu \mu^a_{b+1} ... \}]} 
\Bigr) \nn
&& ~~~~~~~~~~~~~~~~~~~~~~~~~~~~~~~~~~~
 + ( \nabla_\mu J^{[q,\{ \mu^i_j \}]} )
 \pi_{[q,\{ \mu^i_j \}]} 
\Bigr].
\eqa
It is noted that $J^m$'s and $\pi_m$'s are 
$D$-dimensional contra-variant tensors with weight zero
and covariant tensor density with weight one, respectively.
To obtain the `Hamiltonian constraint' ${\cal H}$,
one has to convert Eq. (\ref{pi1}) 
to express $P_{[2,\mu\nu]}$ as a linear combination of $\pi_m$'s
and plug in the expression to Eq. (\ref{sh}).
Since the full expression is complicated, 
we focus on the metric and its conjugate momentum.
Among many other terms, ${\cal H}$ includes 
the linear and quadratic terms for the conjugate momentum,
the cosmological constant and
the $D$-dimensional curvature,
\bqa
{\cal H} & = &
\td A^{\mu \nu}[J(x)] \pi_{[2,\mu \nu]}
- \frac{ \td B^{\mu \nu \lambda \sigma }[ J(x) ]}{  \sqrt{ |G| }     } 
\pi_{[2,\mu \nu]} \pi_{[2,\lambda \sigma]} \nn
&&
- \sqrt{|G|}
\Bigl\{
C_0[J(x)] + C_1[J(x)] {\cal R} 
\Bigr\} + ..., 
\label{sh2}
\eqa
where $...$ represents the higher dimensional terms 
that involve covariant derivatives of $\pi$ and the curvature.
Cubic or higher order terms in $\pi_{[2,\mu\nu]}$ are not allowed
because at most double-trace operators are generated 
out of single-trace operators at each step of RG.
The linear term in the conjugate momentum 
arises because the operators that are quartic in $\Phi$,
such as $\frac{1}{N} \tr[ \Phi^3 \nabla_\mu \nabla_\nu \Phi]$,
renormalizes the metric through the quantum correction 
in Eq. (\ref{c1}).
It is interesting to note that the kinetic term for the conjugate momentum
originates from the beta function under the RG,
while the potential term for the metric originates from 
the Casimir energy.
Besides the dynamical gravitational mode, 
the theory also includes other degrees of freedom,
including the higher spin fields 
for $\frac{1}{N} \tr [ \Phi \nabla_{\mu_1} 
 \nabla_{\mu_2} ... 
 \nabla_{\mu_n} \Phi]$
and the fields associated with the single-trace operators 
that are cubic or
higher order in $\Phi$ .
As was noted in Sec. IV, the latter fields 
do not have the bare `potential energy'
because the Casimir energy is independent of those fields.
However, they do have the quadratic kinetic term in general
because double-trace terms are generated for those operators under the the RG.
Although the bare action for those higher order sources are ultra-local
along the $D$-dimensional space, 
potential terms that involve derivatives along the $D$-dimensional space
will be generated dynamically, 
as other heavier fields are integrated out in the bulk\cite{SLEE112}.

In the large $N$ limit, the bulk fields become classical.
In particular, non-perturbative fluctuations of the bulk fields
are dynamically suppressed\cite{SLEE112}.
The on-shell action in the bulk
computes the partition function 
of the original matrix field theory
in the large $N$ limit.
The classical equation of motion is given by
\bqa
\partial_z J^n  =  \{ J^n, {\bf H} \}, ~~~
\partial_z \pi_n  =  \{ \pi_n, {\bf H} \}, 
\label{EOM}
\eqa
where the Poisson bracket is defined by
\bqa
\{ A, B \} & = & \int d^{D} x  \left[
\frac{ \delta A}{\delta J^n}
\frac{ \delta B}{\delta \pi_n}
-
\frac{ \delta A}{\delta \pi_n}
\frac{ \delta B}{\delta J^n}
\right].
\eqa
To solve the equation of motion, we need another set of boundary conditions besides $J^n(x,0) = {\cal J}^n(x)$.
The second set of boundary condition is dynamically imposed by the regularity condition in the IR limit\cite{SLEE10}.
At the saddle point, we have
\bqa
Z[{\cal J}(x)] 
=
e^{i (\bar S_{UV} + \bar S)}
Z[J(x,\infty)], 
\label{Z2}
\eqa
where the bulk action is evaluated at the saddle point configuration,
and $J(x,\infty)$ is determined from the condition that the bulk action is finite in the IR limit.
At the first glance, this expression does not seem meaningful
because both $Z[{\cal J}(x)]$
and $Z[J(x,\infty)]$ are not well defined
due to the divergent determinants.
However, correlation functions
are well defined because the 
divergences in the determinants cancel.
Correlation functions of local operators are given by
\bqa
\left< O_{n_1} (x_1) O_{n_2} (x_2) ... O_{n_k} (x_{n_k}) \right>
= \left.
\frac{i^n}{N^{2n} Z}\frac{ \partial^k Z[{\cal J}^{'}(x)] }
{ \partial w^{n_1} \partial w^{n_2} .. \partial w^{n_k} }
\right|_{w=0}
\label{cor}
\eqa
where
\bqa
{\cal J}^{'n}(x)  = {\cal J}^{n}(x)
+  \sum_{i=1}^k w^{n_i} \delta^{n}_{ n_i} \delta(x-x_i). 
\label{per}
\eqa
If non-local sources are turned on, 
both the saddle point solution in the bulk
and $J(x,\infty)$ are modified.
For perturbations that are localized both in space and time in $D$-dimensions
as the one in Eq. (\ref{per}),
it is expected that $J(x,\infty)$ does not depend on the perturbation, $w^{n_i}$.
This is because local perturbations are always 
irrelevant and die out in the IR limit in unitary theories.
Therefore, the contribution 
from $Z[J(x,\infty)]$ will drop out in Eq. (\ref{cor}).

It is noted that the $D$-dimensional general covariance
does not allow mass term for the metric.
This does not preclude the possibility that metric fluctuations
become massive as other fields are condensed, 
breaking the $D$-dimensional Lorentz symmetry.
Here we consider the simple case where the sources
${\cal J}^m(x)$ at the boundary  respect the $D$-dimensional 
Lorentz symmetry, and 
the vacuum does not break the symmetry spontaneously.
In this case, the saddle point configuration of the bulk fields 
will also respect the $D$-dimensional Lorentz symmetry.
Mathematically, this means that the only tensor that has
a non-zero spin and has a non-zero expectation in the bulk 
is the metric.
Then, the tensors for the conjugate momentum in Eq. (\ref{sh2}) 
take the form,
\bqa
\td A^{\mu \nu}[J(x)] &=& A[J(x)] G^{\mu \nu}, \nn
\td B^{\mu \nu \lambda \sigma }[ J(x) ] &=& 
 B_1[ J(x) ] G^{\mu \nu} G^{\lambda \sigma}
+ B_2[ J(x) ]
 G^{\mu \lambda} G^{\nu \sigma}
\eqa
at the saddle point, where 
$A[J(x)]$, 
$B_1[ J(x) ]$
and $B_2[ J(x) ]$ are
scalar functions of the sources and their covariant derivatives. 

Although the theory in the bulk is a quantum theory of 
dynamical metric in $(D+1)$-dimensional space,
it is not clear whether this theory 
has the diffeomorphism invariance in the bulk, 
which is the key property of gravitational theories.
In the canonical formalism,
the $(D+1)$-dimensional diffeomorphism invariance 
would show up as $(D+1)$ first-class constraints.
If $A[J(x)]$, $B_1[ J(x) ]$ and $B_2[ J(x) ]$ were just constants,
they have to satisfy specific conditions in order for 
the Hamiltonian constraint to be first-class.
For generic values of $A$, $B_1$ and $B_2$, 
the Hamiltonian constraint ${\cal H}$ is not first-class,
in which case the theory does not have the full
$(D+1)$-dimensional diffeomorphism invariance.
Given that the coefficients are dynamically determined, 
it seems highly unlikely that they have the saddle point values
of the fixed ratio at all points in the bulk 
independent of ${\cal J}^n$.
However, we have to be more careful here because 
the present theory is not a pure gravitational theory.
As a result,
$A$, $B_1$ and $B_2$
depend on other dynamical fields which themselves have non-trivial Poisson bracket
with their own conjugate momenta.
Namely, we can not just replace $A$, $B_1$ and $B_2$ with the saddle point values
when we determine the nature of the constraint.
In other words, one should compute the Poisson bracket 
among the constraints, treating all dynamical fields on the equal footing. 
Instead of computing the Poisson bracket explicitly,
here we use a simple argument to show that all $(D+1)$-constraints
are first-class.

As was emphasized in Secs. IV and V, 
the partition function does not depend on the choice of the lapse
$N^D(x,z) = \alpha(x,z)$ and the shift $N^\mu(x,z)$.
From the fact that the partition function is independent of $N^M(x,z)$, we obtain
\bqa
< {\cal H}_M(x,z) > = \frac{1}{Z} \frac{\delta Z}{\delta N^M(x,z)} =0.
\eqa
Therefore the lapse and the shift play the role of Lagrangian multipliers 
which impose the local constraints,
\bqa
{\cal H}=0, ~~~{\cal H}_\mu=0
\eqa
inside the bulk spacetime.
Since the above equality holds at any time $z$, 
we have
\bqa
\frac{\partial}{\partial z} < {\cal H}_M(x,z) > = 
\int d^Dy ~ N^{M^{'}}(y,z) 
\left< \{ {\cal H}_M(x,z), {\cal H}_{M^{'}}(y,z) \} \right> 
= 0. 
\eqa
In order for this to be true for any choices of $N^M(x,z)$, 
we have
\bqa
\{ {\cal H}_M(x,z), {\cal H}_{M^{'}}(y,z) \} = 0
\eqa
at the saddle point.
This implies that the $(D+1)$ constraints are first-class classically.
These constraints generate local spacetime transformations in the bulk.
The Hamiltonian constraint ${\cal H}$ generates the transformation,
\bqa
x^\mu \rightarrow x^\mu, 
&& 
z \rightarrow z + dl N^D(x,z),
\eqa 
whereas the momentum constraint ${\cal H}_\mu$ generates
\bqa
x^\mu \rightarrow x^\mu + dl N^\mu(x,z), 
&& 
z \rightarrow z.
\eqa 
A general $(D+1)$-dimensional diffeomorphism generated by a combination of the two
corresponds to choosing a different prescription for the local RG procedure.

\section{A simple example}

Because the construction is rather complicated,
it will be useful to apply the prescription to a simple toy model
to illustrate the backbone idea.
Here we provide an explicit construction for the simplest possible
matrix model : $0$-dimensional matrix theory.
The partition function is given by
\bqa
Z[ {\cal J}] & = & \int d \Phi ~~ \exp \left[ 
i \left( - N \sum_{n=2}^\infty {\cal J}^n \tr( \Phi^n )  + N^2 V[ \tr( \Phi^n )/N ]  \right) \right],
\label{Ze}
\eqa
where 
$\Phi$ is a real traceless symmetric matrix 
and
$V[ \tr( \Phi^n )/N ]$ is a general non-linear function of single-trace operators 
which may be expanded as
\bqa
V = \sum_{q=2}^\infty \sum_{m_1,m_2,...,m_q} N^{-q} {\cal J}^{m_1,m_2,..,m_q} \tr( \Phi^{m_1} ) \tr( \Phi^{m_2} ) .. \tr( \Phi^{m_q} ).
\eqa
Because there is no spacetime,
the partition function is given by a single matrix integration.
${\cal J}^n$'s (${\cal J}^{m_1,m_2,..,m_q}$'s) represent the sources for 
the single-trace (multi-trace) operators.
We assume that the sources have small imaginary components
so that the integration is well defined, e.g., 
$Im {\cal J}^n = -\epsilon$ for even $n$;
$Im {\cal J}^n = 0$ for odd $n$.

To remove the multi-trace operators in $V$,
we introduce a pair of auxiliary fields,
$  J^{(1), n}$, $P^{(1)}_n$
 for each single-trace operator, 
\bqa
Z & = & \int d  J^{(1), n} d P^{(1)}_n  d  \Phi ~~ 
e^{i S_1 },
\eqa
where
\bqa
S_1 & = & N^2 \left\{  J^{(1), n } \left( P_{n}^{(1)} -  \frac{\tr( \Phi^n )}{N} \right) - {\cal J}^{n} P_{n }^{(1)}  + V[ P_{m}^{(1)} ] \right\}
\label{eL1}
\eqa
with $V[ P_{m}^{(1)} ] = \sum_{q=2}^\infty \sum_{m_1,m_2,...,m_q} {\cal J}^{m_1,m_2,..,m_q} 
P_{m_1}^{(1)} P_{m_2}^{(1)} .. P_{m_q}^{(1)}$.
The contours of $P^{(1)}_n$'s are along the real axis,
but the contours of $J^{(1),n}$ are chosen slightly off the real-axis
as $Im J^{(1),n} = Im {\cal J}^n$,
which guarantees that integration for $\Phi$ is well defined.
Now we have only single-trace operators for $\Phi$
which are coupled to the dynamical sources $J^{(1),n}$.
$P^{(1)}_n$ is the conjugate variable which corresponds to the single-trace operator, 
$\frac{1}{N} \tr ( \Phi^n )$.
This can be seen from Eq. (\ref{eL1}) where
$J^{(1),n}$ plays the role of a Lagrangian multiplier
which enforces the constraint, $P^{(1)}_n = \frac{1}{N} \tr ( \Phi^n )$.

There is only one operator which is quadratic in $\Phi$.
We use its source $J^{(1),2}$ as a scale to generate a renormalization group transformation.
Because $< \Phi^2 > \sim 1/ J^{(1),2}$, 
we can regard $1/J^{(1),2}$ as a UV cut-off, 
and generate RG flow by lowering $1/J^{(1),2}$\cite{SLEE10}.
The fact that $1/J^{(1),2}$ plays the role of a UV cut-off
can be understood from the observation that 
with a smaller $1/J^{(1),2}$
the fluctuations of $\Phi^2$ decreases.
Using the method described in Eq. (\ref{NB}), the original matrix field $\Phi$ can be written
as a sum of the low energy field $\phi$ and the high energy field $\td \phi$,
\bqa
Z & = & 
\left[ \frac{ \td m^2  J^{(1),2'}}{J^{(1),2}} \right]^{\frac{(N+2)(N-1)}{4}}
\int 
d  J^{(1), n} d P^{(1)}_n
d \phi d \td \phi ~ 
e^{i S_2},
\eqa
where
\bqa
S_2 & = & 
 N^2 \Bigl\{  
V[ P_{m}^{(1)} ] + ( J^{(1), n } - {\cal J}^n )  P_{n}^{(1)} \Bigr\} \nn
&& 
- N  J^{(1),2'} \tr ( \phi^2 ) - N \td m^2 \tr ( \td \phi^2 )
- N \sum_{n=3}^\infty J^{(1),n} \tr (\phi+\td \phi)^n
\eqa
with $ J^{(1),2'} = J^{(1),2} e^{2 \alpha^{(1)} dz}$ and $\td m^2 = \frac{J^{(1),2}}{2 \alpha^{(1)} dz}$.
Here $dz$ is an infinitesimal parameter and $\alpha^{(1)}$ is the rate
at which the UV cut-off is lowered in the first step of RG.
Integrating out the high energy mode,
one obtains the effective action 
which includes
a Casimir energy,
quantum corrections to the single-trace operators, 
and double-trace operators,
\bqa
S_3 & = & 
 N^2 \Bigl\{  
V[ P_{m}^{(1)} ] + ( J^{(1), n } - {\cal J}^n )  P_{n}^{(1)} \Bigr\}  
-i \alpha^{(1)} dz \frac{ (N+2)(N-1)}{2} \nn
&&
- N \left( J^{(1),n} + \alpha^{(1)} dz A^n[J^{(1)}] \right) \tr (\phi^n)
+ \alpha^{(1)} dz B^{nl}[ J^{(1)} ]  \tr (\phi^n) \tr (\phi^l),
\eqa
where
\bqa
A^n[J^{(1)}] & = & 2 J^{(1),2} \delta_{n,2} + \frac{1}{2 J^{(1),2}}
\left[
\sum_{k+l = 2+n} l k J^{(1),k} J^{(1),l} + \frac{i (n+1)(n+2) }{2 N}\left( 1-\frac{2}{N} \right) J^{(1), n+2}
\right], \nn
B^{nl}[J^{(1)}] & = &  \frac{1}{2 J^{(1),2}}
\left[
(l+1) (n+1) J^{(1), l+1} J^{(1), n+1} - \frac{ i (n+l+2)}{2} J^{(1), n+l+2}
\right].
\eqa
In this $0$-dimensional matrix model, the Casimir energy
is a constant independent of the sources.
However, in higher dimensions, the Casimir energy is in general 
a function of dynamical sources, including metric as we saw in Sec. IV.
Now we introduce another set of auxiliary fields $J^{(2),n}$, $P^{(2)}_n$
to remove the double-trace operators that are generated from quantum corrections.
Then $\phi$ is again divided into the low energy mode and the high energy mode,
by rescaling $J^{(2),2}$ by $e^{2 \alpha^{(2)} dz}$.
Integrating out the high energy mode generates 
double-trace operators, which are removed by another set of auxiliary fields.
Repeating these steps, one can write down the original partition function
in terms of the integration of the dynamical sources $J^{(k),n}$
and the conjugate variables $P^{(k)}_n$ introduced at each step of RG.
As $dz \rightarrow 0$, 
the discrete RG step becomes a continuous dimension,
and $J^{(k),n}$, $P^{(k)}_n$, $\alpha^{(k)}$ become functions of $z$ :
$J^n(z)$, $P_n(z)$, $\alpha(z)$.
The original $0$-dimensional theory in Eq. (\ref{Ze}) 
is now mapped into an one-dimensional theory,
\bqa
Z[ {\cal J} ] & = & \left. 
\int {\cal D}J^n(z) {\cal D}P_n(z) ~~ 
e^{i \Bigl( S_{UV}[P(0)] + S[J(z),P(z)] + S_{IR}[J(T)] \Bigr) } \right|_{J^n(0) = {\cal J}^n}, 
\label{Ze2}
\nn
\eqa
where
\bqa
S_{UV} & = & N^2 V[ P_m(0) ], \nn
S & = &  N^2 \int_0^T dz ~~ 
\left[
 ( \partial_z J^n )  P_n - \alpha(z) {\cal H} \right], \nn
S_{IR} & = & -i \ln \int d \phi ~~ e^{ -i N \sum_n J^{n}(T) \tr ( \phi^n ) }.
\eqa
Here $T$ is the RG \lq{time}\rq{} at which we stop the coarse graining procedure.
This creates a boundary at $z=T$ and a boundary action $S_{IR}$.
The partition function is independent of $T$, 
and one can take $T \rightarrow \infty$ 
to push the `IR boundary\rq{} to infinity.
The `Hamiltonian\rq{} in the bulk is given by
\bqa
{\cal H} & = &
i \frac{ (N+2)(N-1)}{2 N^2} 
+ 2 J^2 P_2 \nn
&& + \frac{1}{2 J^{2}} \sum_{n \geq 2}
\left[
\sum_{k,l \geq 3; k+l = 2+n} l k J^{k} J^{l} + \frac{i (n+1)(n+2) }{2 N}\left( 1-\frac{2}{N} \right) J^{n+2}
\right] P_n \nn
&& - \frac{1}{2 J^{2}}
\left[
\sum_{l,n \geq 2} (l+1) (n+1) J^{l+1} J^{n+1} 
P_n P_l
- 
\sum_{l+n \geq 2}
\frac{ i (n+l+2)}{2} J^{n+l+2}
P_n P_l
\right].
\eqa
For $P_n$ with $n<2$ we use the convention, $P_0=1$ and $P_1=0$,
which reflect the fact that $\frac{1}{N} \tr(I)=1$ and $\frac{1}{N} \tr(\phi)=0$.
This is the one-dimensional holographic description for the $0$-dimensional matrix theory.

From the expression in Eq. (\ref{Ze2}),
one immediately realizes that the partition function can be viewed 
as a transition amplitude of a quantum mechanical system 
if $z$ is identified as time and ${\cal H}$ as Hamiltonian.
In the Hamiltonian interpretation, 
the sources and the conjugate fields are promoted to quantum operators,
and satisfy the commutation relation,
\bqa
[ \hat J^n, \hat P_m ] = i \frac{1}{N^2} \delta^n_m,
\eqa
where 
$[\hat A, \hat B] = \hat A \hat B - \hat B \hat A$ is the usual commutator
and $\frac{1}{N^2}$ plays the role of the Planck constant.
The partition function can be written as
\bqa
Z & = & < \Psi_f | e^{ - i N^2 \int_0^T dz \alpha(z) {\cal H} } | \Psi_i >,
\eqa
where the initial and the final wavefunctions are given by
\bqa
<P_n|\Psi_i > & = & e^{ -i N^2 {\cal J}^n P_n + i S_{UV}[P_n] }, \nn
<J^n|\Psi_f >^* & = & e^{ i S_{IR}[J^n] }.
\eqa
Since Hamiltonian is not Hermitian, the evolution is not unitary,
which is consistent with the fact RG flow is irreversible.
Note that $\alpha(z)$ becomes the lapse function along the time direction.
Moreover, the partition function is independent of the choice of $\alpha(z)$.
Choosing a different $\alpha(z)$ amounts to using a different parameterization along the RG flow.
The independence of the partition function under the reparameterizaion of RG flow 
is nothing but the diffeomorphism invariance of the bulk theory.
This one-dimensional diffeomorphism invariance is expressed in terms of the constraint,
\bqa
\frac{ \delta Z}{\delta \alpha(z)} = -i N^2 < \Psi_f | ~e^{-iN^2 \int_z^T \alpha(w) {\cal H} dw } ~{\cal H} ~ e^{-iN^2 \int_0^z \alpha(w) {\cal H} dw } ~|\Psi_i> = 0.
\eqa
However, ${\cal H} | \Psi_i > $ does not identically vanish 
because the boundary at UV explicitly breaks the diffeomorphism invariance.

\section{Holographic (quantum) RG vs. Conventional (classical) RG}

\begin{figure}[h!]
\centering
      \includegraphics[height=6cm,width=14cm]{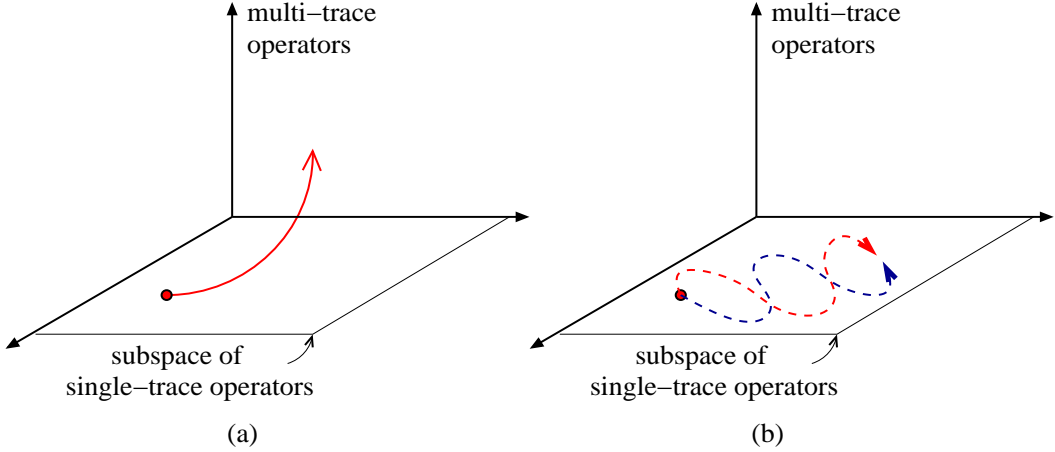}
\caption{
(a) In the conventional RG, multi-trace operators are generated at low energies
although only single-trace operators are turned on at UV. 
Once the initial condition is given,
there is a unique RG trajectory determined by the {\it classical} beta function.
(b) In the holographic description, one only needs to keep track of single-trace operators
under the RG flow at the expense of making the sources for the single-trace operators dynamical variables.
The partition function is given by sum over all possible RG trajectories 
for the single-trace operators.
One has the freedom to employ different RG schemes, namely, 
one can choose different `speed' of RG flow at different scales.
This freedom corresponds to the diffeomorphism invariance in the bulk.
}
\label{fig:4}
\end{figure}

Finally, we compare the holographic description with the conventional RG.
In the present construction, 
the effective action contains 
only single-trace operators at all energy scales.
This greatly simplifies the RG procedure.
The price one has to pay is that one has to promote 
the sources of the single-trace operators to dynamical variables.
In other words, the couplings are not mere constants any more, 
but they have non-trivial quantum fluctuations.
Accordingly, the beta functions that govern the change of couplings 
under the RG flow are {\it quantum} operator equations not classical equations,
\bqa
\frac{i}{N^2} \frac{\partial \hat J^n}{\partial z} &=& [ \hat J^n, \hat {\cal H} ], 
\eqa
where  we have chosen the gauge $\alpha(z) = 1$.
The equation for the sources has to be supplemented 
by the equation for the conjugate operators, 
$\frac{i}{N^2} \frac{\partial \hat P_n}{\partial z} = [ \hat P_n, \hat {\cal H} ]$. 
In the large $N$ limit, the quantum beta function reduces to the classical equation in Eq. (\ref{EOM})
with the identification $ [A,B] \rightarrow \frac{i}{N^2} \{ A, B \}$
where $\{ A, B \}$ is the Poisson bracket.
The `Hamiltonian' that governs the {\it quantum} RG flow is 
dynamically generated from integrating out high energy modes at each step of RG.
This is in contrast to the conventional RG where
the RG trajectory is uniquely determined once an initial condition is given.
In conventional RG,
there is no quantum fluctuations for coupling constants, 
but one has to keep all multi-trace operators along the RG flow.
This is illustrated in Fig. \ref{fig:4}.

\section{Summary and Discussions}
From a first-principle construction, 
it is shown that 
a $D$-dimensional matrix field theory is mapped into 
a $(D+1)$-dimensional quantum theory of gravity,
where the metric in the bulk spacetime is fully dynamical.
The construction starts from the observation that 
one can identify high energy modes as 
fluctuating sources for the low energy modes 
in RG\cite{SLEE10}.
For matrix field theories, this is implemented by
introducing a dynamical source and its conjugate momentum
for each primary single-trace operator 
to remove multi-trace operators at each step of RG\cite{SLEE112}.
In particular, there is a spin two source and its conjugate momentum
that represent the dynamical metric and the energy-momentum tensor, respectively.
While the dynamical sources and momenta are initially
introduced as auxiliary fields, they acquire non-trivial dynamics
as high energy modes are integrated out.
On the one hand, the double-trace operators that are generated 
from single-trace operators 
through quantum correction provides the quadratic kinetic 
term for the conjugate momenta.
On the other hand, the potential terms,
including the curvature term for the $D$-dimensional 
dynamical metric,
are generated from the source dependent determinant for
the high energy mode that is integrated out at each step of RG.
The kinetic and potential terms together can be viewed
as a $(D+1)$-dimensional action written 
in the canonical formalism,
once the extra dimension corresponding to the length scale of RG
is interpreted as a time.
The bulk theory takes the form of quantum theory of gravity
coupled with matter fields of various spins.
Because of the freedom to choose different local RG scheme
without modifying the partition function,
one has $(D+1)$-dimensional diffeomorphism in the bulk,
which in turn leads to $(D+1)$ local constraints.
The Hamiltonian constraint originates from the gauge freedom 
in choosing the spacetime dependent speed of coarse graining in the local RG procedure, 
while the $D$ momentum constraints are associated with relabeling the $D$-dimensional coordinates of low energy modes relative to the coordinates of the high energy modes.
Because different choices of local RG scheme merely correspond 
to choosing different gauge, the $(D+1)$ local constraints are first-class.

The holographic dual for the matrix model includes dynamical gravity
and other fields.
Generically, the cosmological constant in the bulk
is expected to be order of the UV cut-off of the boundary field theory.
Then the saddle point geometry in the bulk will have a curvature 
that is comparable to the scale that controls the gradient expansion 
for the action in the bulk.
In this case, there will be no sense of locality
within the distance scale over which the bulk spacetime is flat,
although the geometry is classical due to the suppressed quantum fluctuations
for a sufficiently large $N$.
It would be of great interest to find boundary field theories
whose gravity dual have a  weakly curved bulk spacetime
through the explicit construction.
This would require stabilizing the theory at a strong coupling.

For a general field theory,
it is not easy to derive the dual theory in a closed form because
one has to keep a large number of fields in the bulk.
However, we have a concrete prescription to identify gravitational duals
starting from boundary field theories.
Using this prescription, one can try to examine the properties of the field theories
which have simple gravity duals.
For example, it will be interesting to see if one can identify the field theory
whose holographic dual is the pure gravity.

\section{Acknowledgment}

 I would like to thank
 Laurent Freidel,
Sean Hartnoll,
Subir Sachdev,
Andrew Tolley,  
Xiao-Gang Wen,
and 
 those who attended 
 the quantum gravity group meeting
 at the Perimeter Institute 
 where a version of this work was presented
for helpful comments. 
This research was supported in part by 
the Natural Sciences and Engineering Research Council of Canada
and the Early Research Award from the Ontario Ministry of Research and Innovation.
Research at the Perimeter Institute is supported 
in part by the Government of Canada 
through Industry Canada, 
and by the Province of Ontario through the
Ministry of Research and Information.

\section{Appendix A : Existence and uniqueness of canonical metric}

We constructively prove the statement 
that there is one and only one metric
in which the quadratic kinetic term has the canonical form
as in Eq. (\ref{bm}) for a given set of sources.
Under a change of the metric used in covariant derivatives, 
only those operators
that are quadratic in $\Phi$ mix with the kinetic term.
So we focus on the quadratic Lagrangian,
\bqa
{\cal L}^{(2)} & = & - \sum_{n=0}^\infty 
j^{\mu_1,..,\mu_n } 
~ \tr \left[ 
\Phi 
\partial_{\mu_1}
..
\partial_{\mu_n}
\Phi 
\right].
\label{qL}
\eqa
The term with $n=1$ can be absorbed into the term with $n=0$ 
via an integration by part, but it is more convenient to keep it for now.
The goal is to find the metric in which the same Lagrangian is written 
in the canonical form,
\bqa
{\cal L}^{(2)} & = & -\sum_{n=0}^\infty 
\sqrt{|G|}
J^{\mu_1,..,\mu_n } 
~
\tr \left[ 
\Phi 
\nabla_{\mu_1}
..
\nabla_{\mu_n}
\Phi 
\right],
\eqa
where $\nabla_\mu$ is the covariant derivative 
associated with the canonical metric that satisfies the condition,
$G^{\mu \nu} = J^{\mu \nu}$.

The canonical metric will be a local functional of the sources $j^{\mu_1,..,\mu_n }(x)$.
Our strategy is to compute the canonical metric using a gradient expansion of the sources.
Suppose that $G^{\mu \nu}_v$ is the metric that coincides
with the canonical metric up to the $v$-th order in derivative.
Namely, $G^{\mu \nu}_v$ is made of the terms 
that have $v$ {\it or less} derivatives of the sources
in the canonical metric.
The exact canonical metric is $G^{\mu \nu}_\infty$.
The Lagrangian in Eq. (\ref{qL}) can be expressed
in terms of the operators constructed with 
the covariant derivative $\nabla_{\mu}^v$ with the metric $G^{\mu \nu}_v$,
\bqa
{\cal L}^{(2)} & = & -\sum_{n=0}^\infty 
j_v^{\mu_1,..,\mu_n } 
~
\tr \left[ 
\Phi 
\nabla_{\mu_1}^v
..
\nabla_{\mu_n}^v
\Phi 
\right].
\eqa
To the zeroth order in derivative, the canonical metric is 
completely determined from $j^{\mu \nu}$,
\bqa
\sqrt{ |G_0| } G^{\mu \nu}_0 & = & j^{\mu \nu}.
\label{g0}
\eqa
It is noted that $G^{\mu \nu}_0$ itself is completely fixed in $D>2$.
In this metric, 
the source for the two derivative operator becomes
\bqa
j_0^{\mu \nu } & = &
\sqrt{|G_0|} G_0^{\mu \nu} 
+ j^{\mu \alpha \beta} \Gamma^{\nu}_{0; \alpha \beta}
+ j^{\alpha \beta \nu} \Gamma^{\mu}_{0; \alpha \beta}
+ j^{\alpha \mu \beta} \Gamma^{\nu}_{0; \alpha \beta}
+ ...,
\eqa
where $\Gamma^{\nu}_{0; \alpha \beta}$ is the Christoffel symbol 
for the metric $G^{\mu \nu}_0$,
and $...$ represents the terms that
have at least two derivatives of the source $j^{\mu \nu}$,
such as 
$j^{\mu \alpha \beta \gamma} 
\Gamma^\delta_{0; \alpha \gamma} \Gamma^\nu_{0;\beta \delta}$ 
and
$ j^{\mu  \alpha \beta \gamma} 
\nabla_\alpha^0 
\Gamma^\nu_{0;\beta \gamma}$.
To the first order in derivative,
the canonical metric is given by 
\bqa
\sqrt{ |G_1| } G^{\mu \nu}_1 & = & 
\sqrt{ G_0 } G^{\mu \nu}_0
+ j^{\mu \alpha \beta} \Gamma^{\nu}_{0; \alpha \beta}
+ j^{\alpha \beta \nu} \Gamma^{\mu}_{0; \alpha \beta}
+ j^{\alpha \mu \beta} \Gamma^{\nu}_{0; \alpha \beta}.
\eqa
Note that the difference between $G_1^{\mu \nu}$ and $G_0^{\mu \nu}$ has 
one derivative in the source.
If we rewrite the Lagrangian using the covariant operators associated with $G^{\mu \nu}_1$,
the source for the two derivative operator 
$\tr[ \Phi \nabla^1_\mu \nabla^1_\nu \Phi]$ 
differs from $\sqrt{ G_1 } G^{\mu \nu}_1$
by terms that have at least two derivatives of the sources.
If one repeats this procedure, 
one can uniquely determine $G^{\mu \nu}_v$
from $G^{\mu \nu}_{v-1}$ by adding terms that have $v$ derivatives of the sources.
This proves that there is a unique canonical metric for a given set of sources.

There is a useful corollary.
Suppose there are two theories which have canonical kinetic terms 
defined on two different curved backgrounds,
\bqa
{\cal L }_a & = & 
- N \sqrt{|G_a|} ~
G_a^{\mu \nu} 
tr \left[ \Phi \nabla_\mu^a \nabla_\nu^a \Phi  \right] 
+ ...,
\eqa
with $a=1,2$.
If  $G_1^{\mu \nu} \neq G_2^{\mu \nu}$,
the two theories are distinct in the following sense.
If the Lagrangians are re-expressed in terms of 
the operators defined on the flat manifold in Eq. (\ref{o}),
the two theories should have different sets of sources.


\begin{thebibliography}{99}
\bibitem{MALDACENA} J. M. Maldacena, Adv. Theor. Math. Phys. {\bf 2} (1998)  231.
\bibitem{GUBSER} S. S. Gubser, I. R. Klebanov and A. M. Polyakov, Phys. Lett. B {\bf 428} (1998) 105.
\bibitem{WITTEN} E. Witten, Adv. Theor. Math. Phys. {\bf 2} (1998) 253.
%
%
\bibitem{EMIL} E. T. Akhmedov, Phys. Lett. B {\bf 442} (1998) 152; E. T. Akhmedov, hep-th/0202055.
\bibitem{DAS} S. R. Das and A. Jevicki, Phys. Rev. D {\bf 68} (2003) 044011.
\bibitem{Gopakumar:2004qb} R.~Gopakumar, Phys.\ Rev.\  D {\bf 70} (2004) 025009; {\it ibid.} {\bf 70} (2004) 025010.
\bibitem{POLCHINSKI09} I. Heemskerk, J. Penedones, J. Polchinski and J. Sully, J. High Energy Phys. {\bf 10} (2009) 079.
\bibitem{SLEE10} S.-S. Lee, Nucl. Phys. B {\bf 832} (2010) 567.
\bibitem{KOCH} R. Koch, A. Jevicki, K. Jin and J. P. Rodrigues, arXiv:1008.0633. 
\bibitem{HEEMSKERK10} I. Heemskerk and J. Polchinski, arXiv:1010.1264.
\bibitem{Faulkner1010} T. Faulkner, H. Liu and M. Rangamani,  arXiv:1010.4036.
\bibitem{Douglas:2010rc} M. Douglas, L. Mazzucato, and S. Razamat, Phys. Rev. D {\bf 83} (2011) 071701.
\bibitem{SUNDRUM} R. Sundrum, arXiv:1106.4501.
\bibitem{SLEE11} S.-S. Lee, Nucl. Phys. B {\bf 851} (2011) 143.
\bibitem{Radicevic} D. Radicevic, J. High. Energy. Phys. {\bf 12} (2011) 023.
\bibitem{SLEE112} S.-S. Lee, arxiv:1108.2253.
\bibitem{VERLINDE} J. de Boer, E. Verlinde and H. Verlinde, J. High Energy Phys. {\bf 08} (2000) 003.
%
\bibitem{LI} M. Li, Nucl. Phys. B {\bf 579} (2000) 525.
\bibitem{Wen_SL} X.-G. Wen, Phys. Rev. B {\bf 65} (2002) 165113.
\bibitem{SAKHAROV} A. D. Sakharov, Sov. Phys. Dokl. {\bf 12}, 1040 (1968). 
\bibitem{POLCHINSKI84} J. Polchinski, Nucl. Phys. B {\bf 231} (1984) 269.
\bibitem{POLONYI} J. Polonyi, arXiv:hep-th/0110026v2.
\bibitem{ADM} R. Arnowitt, S. Deser, and C. Misner, Phys. Rev. {\bf 116}, 1322 (1959).
\end{thebibliography}
\end{document}